\documentclass[12pt]{article}

\usepackage{graphicx,psfrag,epsf}
\usepackage{enumerate}
\usepackage{natbib}
\usepackage{url} 
\usepackage[noblocks]{authblk}
\usepackage[margin=1in]{geometry}

\usepackage[export]{adjustbox}
\usepackage[ruled,vlined]{algorithm2e}
\usepackage{algpseudocode}
\usepackage{amsfonts}
\usepackage{amsmath}
\usepackage{amsthm}
\usepackage{bbold}
\usepackage{xr}
\usepackage{bm, bbm}
\usepackage{booktabs}
\usepackage[export]{adjustbox}
\usepackage{breakcites}
\allowdisplaybreaks
\usepackage{comment}
\usepackage{color}
\usepackage{dsfont}
\usepackage{enumitem}
\usepackage{extarrows}
\usepackage{float}
\usepackage[T1]{fontenc}
\usepackage[utf8]{inputenc} 
\usepackage{lscape}
\usepackage{mathrsfs}
\usepackage{mathtools}
\usepackage{microtype}
\usepackage{multirow}
\usepackage{nicefrac}
\usepackage[labelformat=simple]{subcaption}

\usepackage{url}
\usepackage{wrapfig}
\usepackage[dvipsnames]{xcolor}
\renewcommand{\baselinestretch}{1.5}

\usepackage{amsfonts}
\usepackage{dsfont}
\usepackage{lipsum}  
\usepackage[utf8]{inputenc}
\usepackage[english]{babel}
\usepackage{amsmath}[american, english]
\usepackage{amsthm}

\usepackage{natbib}
\usepackage[T1]{fontenc}
\usepackage{graphicx}
\usepackage[colorlinks=true,linkcolor=blue,citecolor=blue]{hyperref}

\theoremstyle{plain}
\numberwithin{equation}{section}

\newtheorem{lem}{Lemma}
\newtheorem{prop}{Proposition}
\newtheorem{assump}{Assumption}

\theoremstyle{definition}

\theoremstyle{remark}

\newcommand{\bb}{{\boldsymbol b}}
\newcommand{\bc}{{\bf c}}

\newcommand{\bee}{{\bf e}}

\newcommand{\bs}{{\bf s}}

\newcommand{\bfyy}{{\bf y}}
\newcommand{\md}{{\mathcal{D}}}

\newcommand{\bmu}{{\boldsymbol \mu}}
\newcommand{\bxi}{{\boldsymbol \xi}}
\newcommand{\bet}{{\boldsymbol \eta}}
\newcommand{\bbeta}{{\boldsymbol \beta}}
\newcommand{\bGamma}{{\boldsymbol \Gamma}}
\newcommand{\bSigma}{{\boldsymbol \Sigma}}
\newcommand{\bOmega}{{\boldsymbol \Omega}}

\newcommand{\bPhi}{{\boldsymbol \Phi}}
\newcommand{\bphi}{{\boldsymbol \phi}}

\usepackage{setspace}
\usepackage{titling}
\newcommand{\blind}{1} 
\begin{document}

\def\spacingset#1{\renewcommand{\baselinestretch}%
{#1}\small\normalsize} \spacingset{1}

\if1\blind
{
  \baselineskip=28pt \vskip 5mm
    \begin{center} {\LARGE{\bf Bayesian Changepoint Estimation for Spatially Indexed Functional Time Series}}
    \end{center}

    \baselineskip=14pt \vskip 10mm

    \begin{center}\large
    Mengchen Wang\footnote{\baselineskip=12pt Department of Statistics, University of Illinois at Urbana-Champaign},
        Trevor Harris\footnote{\baselineskip=12pt Department of Statistics, Texas A\&M University},
        Bo Li$^1$
    \end{center}
    \baselineskip=19pt \vskip 15mm 
    \vskip 6mm
} \fi

\if0\blind
{
  \bigskip
  \bigskip
  \bigskip
  \begin{center}
    {\LARGE\bf Bayesian Changepoint Estimation for Spatially Indexed Functional Time Series}
    \end{center}
  \medskip
} \fi

\begin{center}
{\large{\bf Abstract}}
\end{center}

We propose a Bayesian hierarchical model to simultaneously estimate mean based changepoints in spatially correlated functional time series. 
Unlike previous methods that assume a shared changepoint at all spatial locations or ignore spatial correlation, our method treats changepoints as a spatial process. This allows our model to respect spatial heterogeneity and exploit spatial correlations to improve estimation. 
Our method is derived from the ubiquitous cumulative sum (CUSUM) statistic that dominates changepoint detection in functional time series. However, instead of directly searching for the maximum of the CUSUM based processes, we build spatially correlated two-piece linear models with appropriate variance structure to locate all changepoints at once.
The proposed linear model approach increases the robustness of our method to variability in the CUSUM process, which, combined with our spatial correlation model, improves changepoint estimation near the edges.
We demonstrate through extensive simulation studies that our method outperforms existing functional changepoint estimators in terms of both estimation accuracy and uncertainty quantification, under either weak and strong spatial correlation, and weak and strong change signals. Finally, we demonstrate our method using a temperature data set and a coronavirus disease 2019 (COVID-19) study.

\baselineskip=14pt

\par\vfill\noindent

\par\medskip\noindent
{\bf Short title}: Functional Changepoint Estimation

\clearpage\pagebreak \pagenumbering{arabic}
\newpage \baselineskip=24pt


    

\section{Introduction}
\label{sec: intro}

In recent years, there has been a considerable renewed interest in changepoint detection and estimation in many fields, including Climate Science \citep{reeves2007review, lund2007changepoint}, Finance and Business \citep{lavielle2007adaptive, taylor2018forecasting}, and traffic analysis \citep{kurt2018bayesian}. The changepoint problem was first studied by \cite{page1954continuous} for independently and normally distributed time series. 
Since then, changepoint literature has grown tremendously. Methods for changepoints in time series have been developed for both at most one change and multiple changepoints.
Vast methodologies are derived based on the cumulative sum (CUSUM) statistic (e.g., \citealp{wald1947sequential, shao2010testing, aue2013structural, fryzlewicz2014multiple}) which was first introduced by \cite{page1954continuous} to detect a shift in the process mean, though other methods have also been proposed (e.g., \citealp{chernoff1964estimating}; \citealp{maceachern2007robust}; \citealp{sundararajan2018nonparametric}). 

With the proliferation of high-frequency data collection and massive data storage in recent years, functional data has become increasingly common and functional data analysis is an increasingly valuable toolkit. For instance, daily temperature data in a specific year can be considered functional data and analyzed using functional data methods. Consequently, functional time series become prevalent and they usually contain more information than a single time series. Following the previous example, daily temperature data over, say 50 years, can be treated as a functional time series which is much more informative than an annual average temperature series with 50 observations. As for univariate time series, changepoint detection and estimation for functional time series have received particular interest owing to the rise of high-dimensional time series.

Within the functional data analysis (FDA) literature, changepoint detection has primarily focused on the scenario of at most one change. \cite{berkes2009detecting} proposed a CUSUM test to detect and estimate changes in the mean of independent functional sequence data. The comprehensive asymptotic properties for their estimation were further studied in \cite{aue2009estimation}. 
Berkes et al.'s test was then extended to weakly dependent functional data by \cite{hormann2010weakly} and to epidemic changes, for which the observed changes will return to baseline at a later time, by \cite{aston2012detecting}. \cite{zhang2011testing} introduced a test for changes in the mean of weakly dependent functional data using self-normalization to alleviate the use of asymptotic control. Later, \cite{sharipov2016sequential} developed a sequential block bootstrap procedure for these methods. Recently,
\cite{aue2018detecting} proposed a fully functional method for finding a change in the mean without losing information due to dimension reduction, thus eliminating restrictions of functional principal component based estimators.
Other methods in multiple changepoint detection for functional time series can be seen in \cite{chiou2019identifying}, \cite{rice2019consistency}, \cite{harris2020scalable} and \cite{li2021bayesian}.

Environmental data often naturally takes the form of spatially indexed functional data.
Again using our temperature data example, if we observe such functional time series at many weather stations in a region, then we have a spatial functional time series. The study for changepoint estimation with spatially indexed functional time series is relatively scant compared to the abundant literature for data not associated with spatial locations. The possible spatial correlation for spatially indexed data presents both challenges and opportunities for such data analysis. 
It is often not straightforward to model and estimate spatial correlation in statistical analysis. However, appropriately taking into account spatial correlation can effectively improve the statistical inference drawn from the spatial data \citep{shand2018spatially}.   
\cite{gromenko2017detection} tackled the changepoint estimation for spatial functional data by assuming a common break time for all functional time series over the spatial domain. They developed a test statistic as a weighted average of the projected CUSUM with the weights defined as the inverse of the covariance matrix of the spatial data. However, the assumption of a common changepoint over the entire spatial domain can be unrealistic when considering functional data over a vast region such as weather data in a state. Other related work on spatial functional data includes a test for the correlation between two different functional data sets observed over the same region \citep{gromenko2012estimation}, a test for the equality of the mean function in two samples of spatial functional data \citep{gromenko2012testing}, and a nonparametric method to estimate the trend as well as evaluate its significance for spatial functional data  \citep{gromenko2013nonparametric}.

\begin{figure}[H]
  \centering
  \includegraphics[width=.4\linewidth]{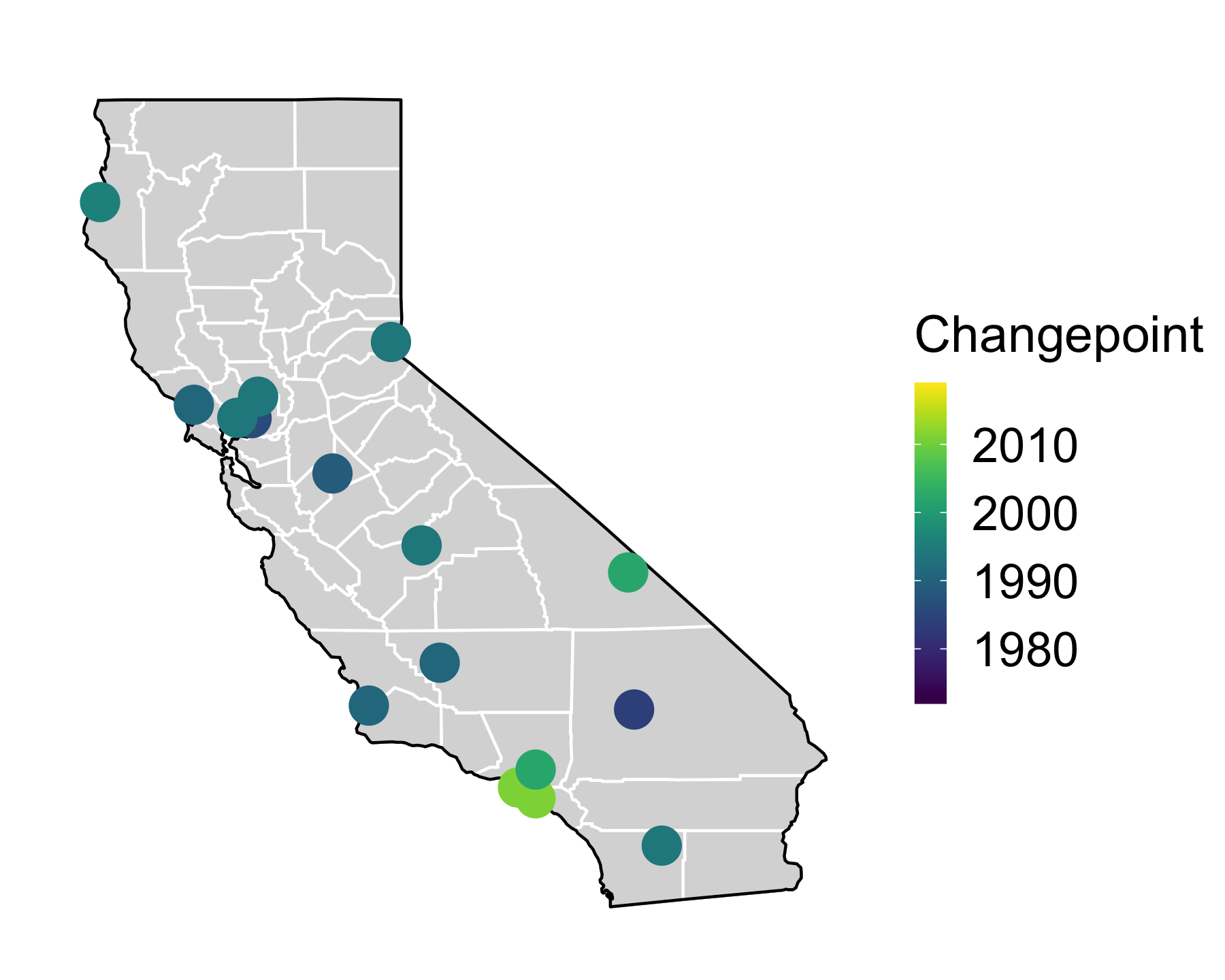}  
  \caption{Changepoint estimates from the FF method at 16 stations. The color of the stations represents the break time of the changepoint. 
  }
  \label{fig: TempCA_FF_cp}
\end{figure}

To illustrate the limitation of assuming a common changepoint for a large region, we examine the changepoints of the daily minimum temperature in California from 1971 to 2020 obtained from \texttt{https://www.ncdc.NOAA.gov/cdo-web/search?datasetid=GHCND}. 
 The data are collected over 207 stations, but only 28 stations have sufficiently complete (<15\% missing values) time series for meaningful change point estimation and are presented here.
 We first use 21 Fourier basis functions to smooth the daily data and then apply the Fully Functional (FF) method of \cite{aue2018detecting} to each station. 
We then test for the existence of changepoints with the FF method and find 16 stations with p < 0.05 after a false discovery rate (FDR) control \citep{benjamini1995controlling}.
The locations of stations and the FF changepoint estimates are shown in Figure \ref{fig: TempCA_FF_cp}. The changepoint estimates appear asynchronous, though somewhat spatially clustered. 
Thus, even without accounting for spatial variability, we can see that the break times vary significantly by location. Assuming just a single common break time would, therefore, misrepresent the changepoint process and lose information.

We propose a flexible changepoint estimation method for simultaneously locating at most one change in each mean function of spatially indexed functional time series. Our method allows both the break time and the amount of change to vary spatially, while taking spatial correlation into account to strengthen the changepoint estimation and respect the inherent spatial continuity. We derive our method based on the asymptotic properties of the functional CUSUM squared norm process at each location. Specifically, we propose to fit spatially correlated piecewise linear models with two pieces for the CUSUM squared norm process across the spatial domain, and estimate changepoints by where the two pieces meet at each individual location. 
All parameters are jointly specified in a Bayesian hierarchical model, which provides a powerful means for parameter estimation as well as allows us to conveniently quantify the uncertainty of the estimation.

The rest of this paper is organized as follows. In Section \ref{sec:methodology}, we first introduce the notations and the properties of the CUSUM squared norm process, then present our proposed method. In Section \ref{sec: simulation}, we conduct simulations under different scenarios to evaluate the performance of our proposed model and other competitive methods. Real data analysis on the California minimum temperatures and the COVID-19 dataset is presented in Section \ref{sec: real}. The paper concludes with a brief discussion in Section \ref{sec: discussion}.

\section{Changepoint Estimation} \label{sec:methodology}
\subsection{Notation and Assumptions}\label{sec:notation}

Let $X_{\bs,t}(u)$ be the functional observation at location $\bs \in \mathcal{D}$ and time $t \in \mathds{Z}$, where $\mathcal{D}$ is a compact subset in $\mathds{R}^d$. Each $X_{\bs,t}(u) \in L^2([0, 1])$ is a real-valued square integrable function defined without loss of generality on the unit interval $[0, 1]$, i.e. $u \in [0, 1]$, and $\int_0^1 \left |X_{\bs,t}^{2}(u) \mathrm{d} u\right|<\infty$.
We assume the functional times series at each location $\bs$ is  generated from the following model,
\begin{equation}
\label{equ: fdmodel}
X_{\bs,t}(u) = \mu_\bs(u) + \delta_\bs(u)\mathds{1}(t > k^*_\bs) + \varepsilon_{\bs,t}(u),\  t \in \mathds{Z},
\end{equation}
where $\mu_{\bs}(u)$ is the baseline mean function that is distorted by the addition of $\delta_\bs(u)$ after the break time $k^*_\bs\in \{1,\dots, T\}$ at location $\bs$, and $\mathds{1}(A)$ is an indicator function that equals $1$ only when event $A$ is true and zero otherwise. We assume that the functional data at all locations are observed at the same time points. To simplify notation, we sometimes suppress $u$ from the functional random variables such as referring to $\varepsilon_{\bs,t}(u)$ by $\varepsilon_{\bs,t}$ when there is no risk of confusion. 

Following Assumption 1 in \cite{aue2018detecting}, we allow the error functions $\varepsilon_{\bs,t}(u) \in L^2([0, 1])$ to be weakly dependent in time by assuming they are $L^p-m-$approximable for some $p>2$. Assumption \ref{assumption} below essentially means that for any location $\bs$ the error functions $\varepsilon_{\bs,t}(u)$ are weakly dependent. 

\begin{assump} 
\label{assumption}
For all spatial locations $\bs \in \md$, the error functions $\left(\varepsilon_{\bs, t}: t \in \mathds{Z}\right)$ satisfy

(a) there is a measurable space $S$ and a measurable function $g: S^\infty \rightarrow L^{2}([0,1])$, where $S^\infty$ is the space of infinite sequences $(\zeta_{t}, \zeta_{t-1}, \ldots)$ with $\left(\zeta_{t}: t \in \mathds{Z}\right)$ taking values in $S$, such that $\varepsilon_{\bs, t}=g\left(\zeta_{t}, \zeta_{t-1}, \ldots\right)$ for $t \in \mathds{Z}$, given a sequence of independent, identically distributed (iid) random variables $\left(\zeta_{t}: t \in \mathds{Z}\right)$; 

(b) there are $m-$dependent sequences $\left(\varepsilon_{t, m}: t \in \mathds{Z}\right)$ such that, for some $p>2$,
$$
\sum_{m=0}^{\infty}\left\{E\left(\left\|\varepsilon_{\bs,t}-\varepsilon_{t, m}\right\|^{p}\right)\right\}^{1 / p}<\infty,
$$
where $\varepsilon_{t, m}=g\left(\zeta_{t}, \ldots, \zeta_{t-m+1}, \zeta_{t, m, t-m}^{*}, \zeta_{t, m, t-m-1}^{*}, \ldots\right)$ with $\zeta_{t, m, j}^{*}$ being independent copies of $\zeta_{0}$ independent of
$\left(\zeta_{t}: t \in \mathds{Z}\right)$.
\end{assump}
This assumption covers most commonly used stationary functional time series models, such as functional auto-regressive and auto-regressive moving average processes. We additionally assume that all error functions are generated from the same distribution as in Assumption \ref{assump: errorid}.
\begin{assump}
\label{assump: errorid}
The errors $\left(\varepsilon_{\bs, t}: \bs \in \mathcal{D}, t \in \mathds{Z}\right)$ are identically distributed random fields on $[0, 1]$.
\end{assump}
Assumption \ref{assump: errorid} indicates that the error functions at all time points and all locations follow the same distribution. Under Model (\ref{equ: fdmodel}), the only changes observed in a functional time series are due to $\delta_\bs(u)$, i.e., changes in the mean of the functional sequence. Therefore, all other aspects of the distribution, such as the variance, are required to remain the same.
While seemingly restrictive, requiring the moments to not change simultaneously  is common in functional time series (\citealp{gromenko2017detection}) and required for identifiablility even in univariate change point estimation (\citealp{horvath1993maximum}).
Practically, Assumption \ref{assump: errorid} also allows to share variance parameters across spatial locations when estimating the properties of error functions. 
Finally, we assume that the error process is stationary and isotropic.
\begin{assump}
\label{assump: 2nd-stationary}
The errors $\left(\varepsilon_{\bs, t}: \bs \in \mathcal{D}, t \in \mathds{Z}\right)$ form a mean zero, second-order stationary and isotropic random field. Formally,
$$
\begin{aligned}
&E\{\varepsilon_{\bs, t}(u)\} = 0,\\
&\operatorname{cov}\{\varepsilon_{\bs, t}(u), \varepsilon_{\bs', t'}(u')\} 
=C(||\bs-\bs'||, t-t', u-u'),
\end{aligned}
$$
where $||\bs-\bs'||$ is the Euclidean distance between spatial locations $\bs$ and $\bs'$.
\end{assump}
Assumption \ref{assump: 2nd-stationary} essentially means the covariance between any two observations only depends on their distance in each dimension, regardless of their locations and relative orientation. 

\subsection{CUSUM Statistic}

Suppose we observe functional time series $X_{s, t}$ at spatial locations $\bs \in \md$
and time points $t= 1, \dots, T$. 
Changepoint detection, at each location $\bs$, can be formulated into the following hypothesis test:
\begin{equation} 
 \label{eqn:hypothesis_test}
     H_0: \delta_\bs=0 \hbox{ versus }  H_A: \delta_\bs \ne 0,
\end{equation}
where $\delta_\bs=0$ means $\delta_\bs (u) = 0,$ for all $u \in [0, 1]$ and otherwise $\delta_\bs \ne 0$. \cite{aue2018detecting} proposed a fully functional approach to testing the hypothesis (\ref{eqn:hypothesis_test}) for each location $\bs$ based on the functional CUSUM defined as
\begin{equation}
\label{equ: cusum def}
S_{\bs, T, k}(u) = \frac{1}{\surd{T}}\left\{\sum_{t=1}^k X_{\bs,t}(u)-\frac{k}{T}\sum_{t=1}^{T} X_{\bs, t}(u)\right\}, \ k=0,\dots, T,
\end{equation}
for which the two empty sums $S_{\bs, T, 0}(u)=S_{\bs, T, T}(u)=0$. 
Noting that the $L^2$ norm of the CUSUM statistic, $\left\|S_{\bs, T, k}\right\|$, as a function of $k$ tends to be large at the true break date motivates
a max-type test statistic for detecting a change in the mean function:
\begin{equation}
\label{eq:cusum-ts}
TS_T(\bs) = \max_{1\leq k\leq T} \|S_{\bs, T, k}(u)\|^2.    
\end{equation}
If a changepoint is detected, \cite{aue2018detecting} further provided an estimator for the break time $k^*_\mathbf{s}$:
$$
\hat{k}^{*}_\bs=\min \left\{k:\left\|S_{\bs, T, k}(u)\right\|=\max _{1 \leq k^{\prime} \leq T}\left\|S_{\bs, T, k^{\prime}}(u)\right\|\right\}.
$$
The CUSUM test based on Equation \eqref{eq:cusum-ts} allows the functional time series to be $m-$ dependent and requires notably weaker assumptions than the functional principal component based methods. The CUSUM statistic is shown to be powerful  (\citealp{page1954continuous}; \citealp{maceachern2007robust}) in detecting mean shift of univariate time series. For functional time series, the CUSUM is also the basis of many other changepoint detection methods \citep{berkes2009detecting,hormann2010weakly, aston2012detecting, sharipov2016sequential, gromenko2017detection}.

\subsection{Properties of Spatial CUSUM Process}

Most previous methods consider 
changepoint detection in a single
functional time series, and thus may have limited power when directly applied for the spatially indexed functional data that exhibit spatial correlation. While \cite{gromenko2017detection} took spatial correlation into account, their assumption of simultaneous changepoint can be too restrictive for data observed in a large spatial domain. We aim to develop a flexible and efficient method to estimate spatially varying break time $k_s^*$ jointly for all locations while taking advantage of spatial correlation in the changepoint estimation. 
Due to the power of CUSUM statistic in changepoint detection, our method will employ the CUSUM as the building block. 

Since our method is derived based on the asymptotic properties of CUSUM processes for spatially indexed functional time series, this section focuses on studying those properties before introducing our model in Section \ref{sec: BHM}. 
To simplify notation, let 
\begin{equation}
Y_{T,k}(\bs)=\left\|S_{\bs, T, k}(u)\right\|^2,  k=0,\dots, T,
\label{equ: CUSUMprocess}
\end{equation}
The notation $Y_{T,k}(\bs)$ emphasises that $Y$ is a spatially varying random process. By definition, $Y_{T,k}(\bs)=0$ when $k=0$ and $k=T$. Since $Y_{T,k}(\bs)$ largely preserves the changepoint information \citep{aue2018detecting}, our method will be built on $Y_{T,k}(\bs)$ which reduces the functional sequence $X_{\bs, t}(u)$ at each location into a time series $Y_{T,k}(\bs)$, $k=0,\ldots,T$. The spatial functional sequence thus reduces into a spatiotemporal random process. 

We then study the characteristics of the spatiotemporal process $Y_{T,k}(\bs)$. Let $\lambda_l$ and $\psi_l(u)$ be the eigenvalues and eigenfunctions of the error process $\epsilon_{\bs,t}(u)$ in Equation (\ref{equ: fdmodel}). The formal definition is deferred to Appendix \ref{app: eigen of kernel}. Let $q=k/T$ be the scaled time point.
\begin{lem}
     \label{thm:m&v_H0}
     Under the null hypothesis of no changepoint at location $\bs$, we have
     \begin{equation}\nonumber
        Y_{T,k} (s)  \overset{\mathcal{D}}{\to}  \sum_{l=1}^\infty \lambda_l B_l^2(q) \hbox{ as } T \to \infty,
    \end{equation}
     where $(B_l: l \in \mathds{N})$ are iid standard Brownian bridges defined on [0, 1],  $E\{\sum_{l=1}^\infty \lambda_l B_l^2(q)\} = q\left(1-q\right)\sum_{l=1}^\infty\lambda_l$ and $\operatorname{var}\{{\sum_{l=1}^\infty \lambda_l B_l^2(q)}
     \}= 2q^2\left(1-q\right)^2\sum_{l=1}^\infty\lambda_l^2$.
\end{lem}

\begin{prop}
     \label{thm:m&v_HA}
     Under the alternative hypothesis that there is one changepoint $k^*_\bs$ at location $\bs$ and the corresponding change function is $\delta_\bs(u)$, we have
    \begin{equation}\nonumber
        \surd\{{Y_{T,k} (\bs)}\} - \surd\{{Z_{T,k}(\bs)}\} \overset{\mathcal{P}}{\to}  0,
    \end{equation}
    for a random process $Z_{T, k}(\bs)$ with 
    \begin{equation}
    \label{eq: HA_mean}
    E\{Z_{T, k}(\bs)\} =
     \begin{cases}
     q\left(1-q\right)\sum_{l=1}^\infty\lambda_l+Tq^2||\delta_\bs(u)||^2\left(1-\frac{k^*_\bs}{T}\right)^2, &\hbox{ if } k\leq k^*_\bs;\\
     q\left(1-q\right)\sum_{l=1}^\infty\lambda_l+T\left(1-q\right)^2||\delta_\bs(u)||^2\left(\frac{k^*_\bs}{T}\right)^2,&\hbox{ if } k >  k^*_\bs,
     \end{cases}
     \end{equation}
    and 
    \begin{equation}
    \label{eq: HA_var}
     \operatorname{var}\{Z_{T, k}(\bs)\} =
     \begin{cases}
     a q^2\left(1-q\right)^2+b_\bs Tq^3\left(1-q\right) \left(1-\frac{k^*_\bs}{T}\right)^2,&\hbox{ if } k\leq k^*_\bs;\\
     a q^2\left(1-q\right)^2+b_\bs T q\left(1-q\right)^3\left(\frac{k^*_\bs}{T}\right)^2,&\hbox{ if } k >  k^*_\bs,
     \end{cases}
     \end{equation}
     where
    $a=2\sum_{l=1}^\infty\lambda_l^2$ and $b_\bs=4\sum_{l=1}^\infty \left\{ \int_0^1 \psi_l(u)\delta_\bs(u) du \right\}^2$.
\end{prop}
Assumption \ref{assump: errorid} for the error functions implies both $\lambda_l$ and $\psi_l$ are invariant across $\bs$ and $t$, so all locations share the same parameter $a$ which represents the feature of the long-run variance, whereas $b_\bs$ depends on change functions that may vary across different locations. Proofs of Lemma \ref{thm:m&v_H0} and Proposition \ref{thm:m&v_HA} are deferred to Appendix \ref{app: proof}.

The asymptotics in Proposition \ref{thm:m&v_HA} indicates that we can use the mean and variance of $\surd\{{Z_{T,k}(\bs)}\}$ to approximate those of $\surd\{{Y_{T,k}(\bs)}\}$ at a large $T$. However, the calculation of the first two moments for $\surd\{{Z_{T,k}(\bs)}\}$ is rather involved compared to that for $Z_{T,k}(\bs)$ due to the square root operator. More details can be found in Appendix \ref{app: proof}. To bypass that difficulty, we propose to use the mean and variance of $Z_{T, k}(\bs)$ to approximate those of the $Y_{T,k}$ process.
This is not an optimal choice, however, we think the approximations are reasonable, at least better than some naive choices such as constant or linearly variance.
To evaluate how well (\ref{eq: HA_mean}) and (\ref{eq: HA_var}) approximate the mean and variance of the $Y_{T, k}(\bs)$ respectively,  we conduct simulations at four different settings composed of two different $T$'s and two signal-to-noise ratio (SNR) values that will be introduced in Section \ref{sec: datageneration}. The details of the simulation can be found in Appendix \ref{app: Y Process}. Figure \ref{fig: mean_var_pairs} compares the empirical mean and variance from the simulations with their theoretical approximations. For all scenarios we considered, the approximations seem to match with the empirical result well, especially in the mean function.   

\begin{figure}[h]
    \centering
    \includegraphics[width=\textwidth]{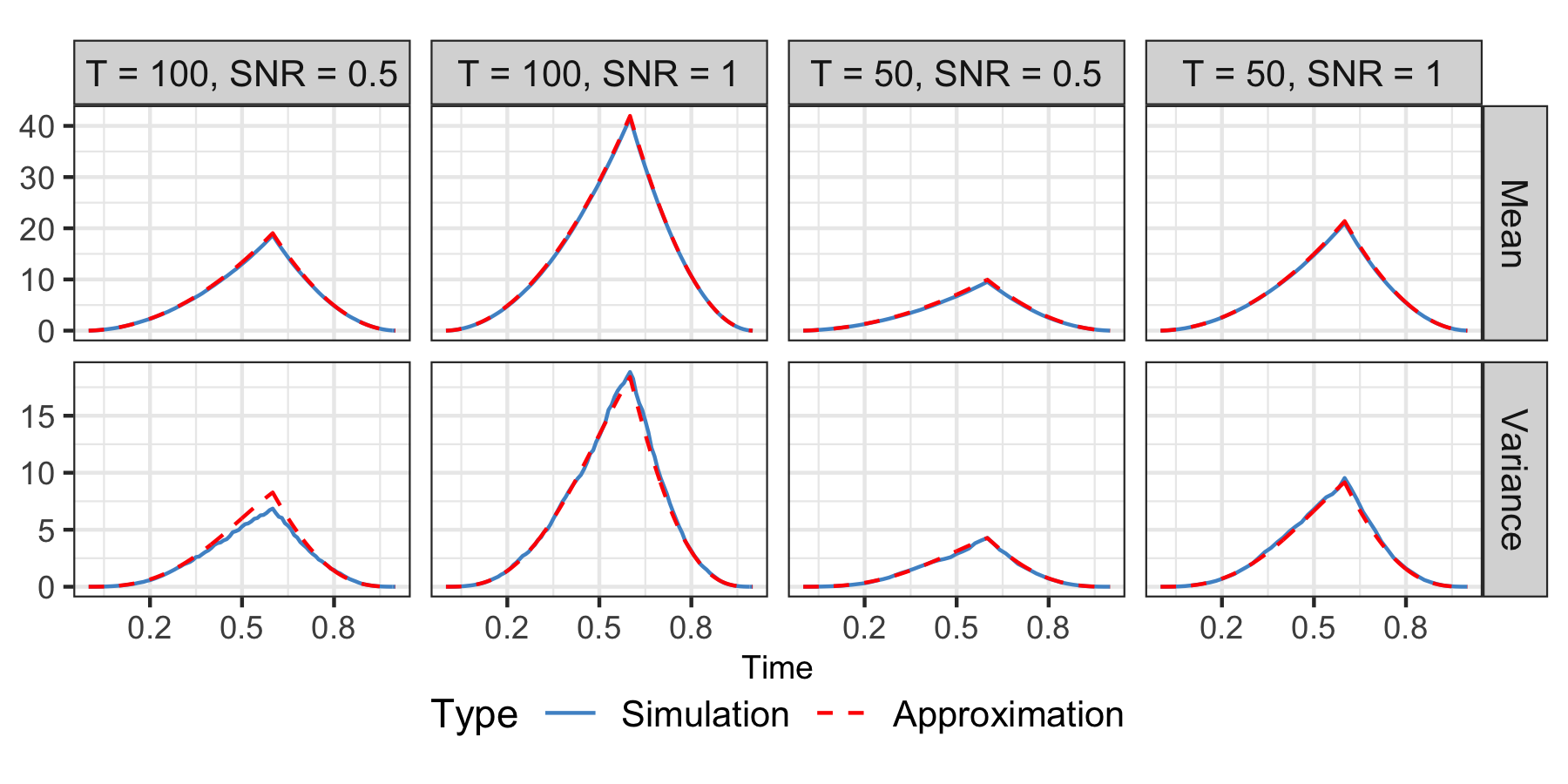}
    \caption{The mean and variance of the $Y_{T,k}$ process based on simulation results (blue solid) and the proposed theoretical approximation (red dashed).}
    \label{fig: mean_var_pairs}
\end{figure}

The expression in  Equation \eqref{eq: HA_mean} shows that when $T$ is large the mean of the $Y_{T,k}(\bs)$ sequence attains its peak at the changepoint.  This is indeed the basis of the test in \cite{aue2018detecting}. 
Figure \ref{fig: mean_var_pairs} also shows that the $Y_{T,k}(\bs)$ sequence starts from exactly zero on both ends and then peaks at the true changepoint 0.6. 
Comparing the mean of $Y_{T,k}(\bs)$ at two different SNR values, it is seen that when the change signal is stronger, the peak tends to be more pointed. The variance of $Y_{T,k}(\bs)$ also starts from zero at the two ends and then increases toward the center. However, there is no theoretical evidence that the variance should maximize at the changepoint. Indeed we find the peak of the variance is not necessarily located at the changepoint, though this particular simulation shows so. 
 
\begin{figure}[h]
    \centering
    \includegraphics[width=0.45\textwidth]{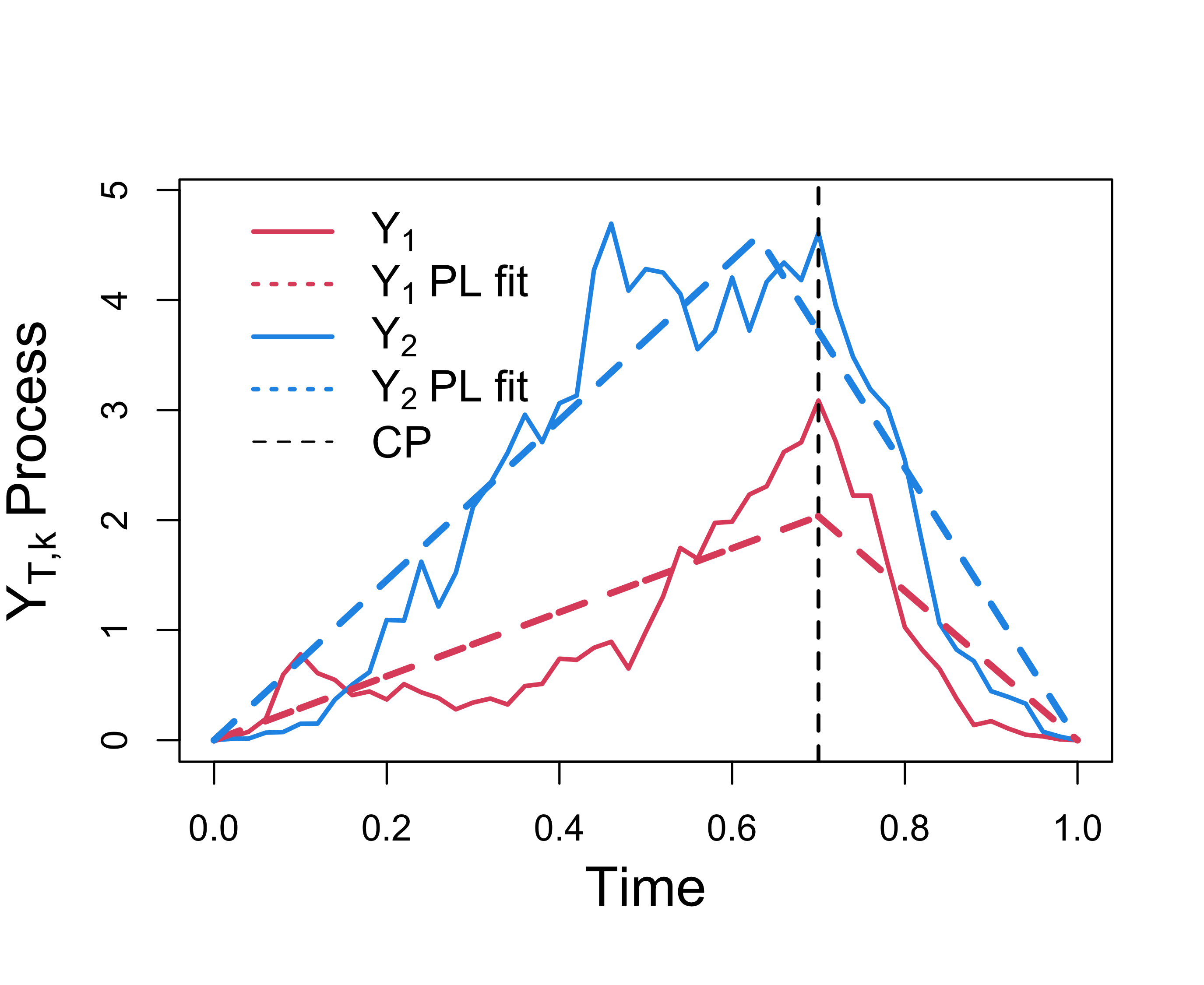}
    \caption{Two simulated $Y_{T,k}$ processes, $Y_1$, $Y_2$ (red and blue solid) with changepoint (CP) at 0.7 (black dashed), and their corresponding fitted two-piece piecewise linear (PL) model (red and blue dashed). Details about the $Y_{T,k}$ process generation and PL fit are in Appendix \ref{app: Y Process}.}
    \label{fig: plm_approx}
\end{figure}

The properties of $Y_{T, k}(\bs)$ enlighten us to estimate the break time by fitting a piecewise linear model with two pieces for the $Y_{T,k}(\bs)$, $0 \le k \le T$ sequence at each location. The two pieces are expected to be joined at the break time. Figure \ref{fig: plm_approx} illustrates this idea using  simulated $Y_{T, k}(\bs)$ processes. Due to the constraint of being zeroes on both ends, the two pieces can be modeled by one slope parameter, and a stronger change signal will lead to a steeper slope. 
Although the mean function in Equation \eqref{eq: HA_mean} suggests a piecewise quadratic model, for simplicity and the robustness of linear models we choose the piecewise linear model which suffices for our purpose of capturing the peak of the $Y_{T,k}$ process. 
In order to correctly quantify the uncertainty of the fitted piecewise linear model and thus the uncertainty of the changepoint estimation, it is important to feed the regression model with the appropriate variance structure. We model the variance of the piecewise linear model following Equation \eqref{eq: HA_var}.

If the functional data are observed at nearby locations, their break times are expected to be similar due to spatial dependency, so is the amount of change. What these similarities pass to the piecewise linear models is that the locations of the joints and the slopes of the models at two neighboring locations tend to be respectively similar. This suggests us to borrow information from neighbors when estimating the changepoint at one specific location. 

Given the above considerations, we propose a Bayesian hierarchical model to jointly estimate changepoints together with their uncertainty for all locations that have changepoints. In practice, we can first apply any changepoint detection method at each location and then employ FDR to adjust the p-values to decide which locations show significant evidence of having a changepoint. If the number of spatial locations $N$ is large, the mirror procedure developed by \cite{yun2020detection} can be an effective alternative to the classic FDR control. 

\subsection{Bayesian Hierarchical Model}\label{sec: BHM}

 We model the $Y_{T,k}(\bs)$ process through a Bayesian hierarchical model. Assume changepoints are detected at locations $\bs_1$, \ldots, $\bs_N$.  
 At each of those locations, we fit a two-piece piecewise linear model with only one slope parameter for $Y_{T,k}(\bs),\ k=1,\ldots, T-1$ due to the constraint of $Y_{T,k}(\bs)=0$ for $k=0$ and $k=T$. We model the slope parameters and the joints of the two pieces as spatially correlated processes to account for the spatial correlation in the break time and change amount of the changepoints. Let $c(\bs)=k^*_\bs/T\in(0,1)$ be the scaled location specific changepoint. We propose the following model: 

\paragraph{Stage I}
 Likelihood of the $Y_{T,k}(\bs)$ process:
$$
Y_{T,k}(\bs) = \beta(\mathbf{s})[\{c(\mathbf{s})-1\}q+\{q-c(\mathbf{s})\}\mathds{1}\{q\geq c(\mathbf{s})\}]+e_k(\mathbf{s}),\ k=1,\dots,T-1,$$
where $\beta(\mathbf{s})<0$ is the spatially varying piecewise linear model coefficient,
and the error process $e_k(\bs)$ is assumed to be a zero-mean spatially correlated Gaussian process. We further assume a space-time separable covariance structure for errors for simplicity, as is widely used in spatiotemporal modeling \citep{haas1995local, hoff2011separable}. We denote the entire error process as
\begin{equation}
\nonumber
    \bee=\left(e_1(\mathbf{s}_1),\dots,\ e_1(\mathbf{s}_N),\ e_2(\mathbf{s}_1),\ldots,e_{2}(\mathbf{s}_N),\ldots,e_{T-1}(\mathbf{s}_1),\ldots, e_{T-1}(\mathbf{s}_N)\right)^T,
\end{equation}
\vspace{-1.5mm}
and assume
\vspace{-1.5mm}
\begin{equation}
\nonumber
    \mathbf{e}\sim N(\textbf{0}_{N(T-1)},\bOmega^{1/2} \bGamma_t \otimes \bGamma_s \bOmega^{1/2}),
\end{equation}
\vspace{-1.5mm}
where
\begin{equation}
\nonumber
    \bOmega = \text{diag}\left(\omega^2_1(\mathbf{s}_1), \ldots,\omega^2_1(\mathbf{s}_N), \ldots,\omega^2_{T-1}(\mathbf{s}_1),\ldots, \omega^2_{T-1}(\mathbf{s}_N)\right),
\end{equation}
with
\begin{equation}
\nonumber
    \begin{aligned}
       \omega^2_k(\mathbf{s}) = 
    \begin{cases}
      a q^2(1-q)^2+b_\bs c(\mathbf{s})^2Tq(1-q)^3, \ &\text{if}\ q>c(\bs);\\
      a q^2(1-q)^2+b_\bs\{1-c(\mathbf{s})\}^2Tq^3(1-q),\   &\text{if}\ q\leq c(\bs),
    \end{cases}       
    \end{aligned}
\end{equation}
and $\mathbf{b}_{d}$ is a $d-$dimensional vector of all $b$ values.
The variance term $\bOmega$ follows the theoretical approximation in Equation (\ref{eq: HA_var}) to represent the uncertainty of $Y_{T,k}(\bs)$. Parameters $a$ and $b_\bs$ are complex functions of unknown eigenvalues, eigenfunctions and change functions. We will  directly treat them as unknown nuisance parameters in our model. This also gives us the leverage of being less dependent on the exact form of the approximation but rather following its basic structure. The pure temporal correlation matrix $\bGamma_t$ and pure spatial correlation matrix $\bGamma_s$ can be governed by any valid correlation function such as exponential or Mat\'ern function \citep{stein2012interpolation}. For simplicity,  we assume an exponential covariance function for both matrices:   
\begin{equation} \nonumber
\bGamma_{t(k,k')}=\exp\left(-\frac{|k-k'|}{T\phi_t}\right), \qquad \bGamma_{s(j,j')}=\exp\left(-\frac{||\mathbf{s}_{j}-\mathbf{s}_{j'}||}{\phi_s}\right),
\end{equation}
where $\phi_t$ and $\phi_s$ are range parameters for temporal and spatial correlation, respectively.

As shown earlier by the asymptotic and numerical results, the shape of the piecewise linear model is influenced by the change function and changepoint. To respect the fact that the nearby locations tend to have similar changepoints and change functions, we regulate $\bbeta = (\beta(\bs_1), \dots, \beta(\bs_N))^T$ and $\bc = (c(\bs_1),\dots,c(\bs_N))^T$ by a correlated process. Since $\bb = (b_{\bs_1},\dots,b_{\bs_N})^T$ also depends on the change function, it is governed by a correlated process as well.  Because the dependency in $\bbeta$, $\bc$ and $\bb$ all arise from the spatial dependency in the data, it is not unreasonable to assume these parameters share one correlation matrix $\bSigma(\phi)$ to retain parsimony of the model. 
Considering the constraints that the slope $\beta(\mathbf{s})$ is negative, changepoint $c(\bs)$ is between 0 and 1, and the parameters $a$ and $b_\bs$ in the variance part are positive, we construct the following priors: 
\paragraph{Stage II} Priors:
    $$
      \begin{aligned}
      \log(-\boldsymbol{\beta})&\sim N(\boldsymbol{\mu}_{\beta}, \sigma_{\beta}^2\bSigma(\phi)),\\
     \Phi^{-1}(\boldsymbol{c})&\sim N(\boldsymbol{\mu}_c, \sigma_c^2\bSigma(\phi)),\\
     \log(a)&\sim N(\mu_a, \sigma^2_a),\\
      \log(\boldsymbol{b})&\sim N(\boldsymbol{\mu}_{b}, \sigma^2_b\bSigma(\phi)),
      \end{aligned}
      $$
     where $$\bSigma(\phi)_{nn'}=\exp\left(-\frac{||\mathbf{s}_{n}-\mathbf{s}_{n'}||}{\phi}\right).$$
    
      All parameters $\bmu_\beta$, $\bmu_c$, $\mu_a$ and $\bmu_b$ take values in $\mathds{R}$, so we choose a normal distribution with large variance as their weak hyperpriors. The variance parameters $\sigma_i^2: i = \beta, c, a, b$ are all given a conjugate inverse gamma hyperprior. We choose $\operatorname{IG}(0.1, 0.1)$ because it provides sufficiently vague hyperpriors for the variances of $\bbeta$, $\boldsymbol{c}$, $a$, and $\bb$. 
      The range parameters $\phi$, $\phi_s$ and $\phi_t$ are positive, so we choose an exponential hyperprior for them but set a different hyperparameter for $\phi_t$, given that the spatial and temporal domains have different characteristics. 
\paragraph{Stage III}
    Hyperprior:
    $$
    \begin{aligned}
       &\boldsymbol{\mu}_i \sim N(\textbf{0}_N, 9\boldsymbol{I}_N),\ i = \beta, c, b,\\
       &\mu_a \sim N(0, 9),\\
       & \sigma^2_i \sim IG(0.1,0.1),\ i = \beta, c, a, b,\\
       &\phi,\ \phi_s \sim \exp(0.5),\\
       &\phi_t \sim \exp(0.1),
    \end{aligned}
      $$
where $\boldsymbol{I}_N$ is the $N \times N$ identity matrix.
We use the Markov chain Monte Carlo (MCMC) algorithm to obtain posterior samples from the model. Gibbs sampling is utilized to sample the posteriors for $\sigma_\beta^2$, $\sigma_c^2$, $\sigma_a^2$ and $\sigma_b^2$, while the Metropolis-Hasting-within-Gibbs algorithm is implemented for the rest parameters. The derivation of posterior distributions can be found in Appendix 
\ref{app: mcmc}.

\section{Simulation Study}
\label{sec: simulation}

We conduct simulations to evaluate the accuracy of our changepoint estimation, as well as the coverage and the length of the credible interval. We also explore how the strength of spatial correlation and change signal influence performance. To further study the properties of our method, we compare it with other competitive methods from the perspective of changepoint estimation.

\subsection{Data Generation}
\label{sec: datageneration}
    
   We randomly select $N=50$ locations in a 10 $\times$ 10 spatial domain as the rejection region $\md_R$ resulting from a changepoint detection algorithm adjusted by the FDR control. Due to the joint estimation for all locations of our method, the false discoveries, i.e., the null locations falsely classified as alternatives, may undermine the estimation. To mimic false discoveries at a typical rate $0.1$, we randomly select a cluster of $N_0=5$ locations among the 50 to be the falsely classified null locations. At each location,  we consider $T=50$ time points and generate $T$ functional data, $X_{\bs, t}(u): u \in [0,1]$ for $t=1,\dots,T$, as defined in Equation (\ref{equ: fdmodel}). At those $N_0$ locations, the change function $\delta_\bs(u)$ is set to be zero. Without loss of generality, we assume the mean curves, $\mu_{\mathbf{s}_1}$, $\dots$, $\mu_{\mathbf{s}_N}$, to be zero functions. Thus, the data generation mainly involves simulating error functions, break time and change functions.  
    
    \paragraph{Error functions:} Although we allow the error functions to be weakly dependent, 
    using temporally independent error functions in simulation studies is very common (\citealp{horvath2013estimation}; \citealp{aue2018detecting}). In particular, \cite{aue2018detecting} repeated their simulation with the first-order functional autoregressive errors, and found the results generally remain the same as those from the independent errors. This is because the $Y_{T,k}(\bs)$ process is insensitive to the error correlation structure. We therefore adopt temporally independent error functions in our simulation. For each location, we generate $T$ error functions $\varepsilon_{\bs, t}$ as follows,
    $$\varepsilon_{\bs, t}(u) = \sum_{l=1}^L \xi _{\bs,t}^l \nu_l(u),\ t=1,\dots,T\ ,\ \bs \in \md_R,$$
    where $L=21$ is the number of Fourier basis functions, $\nu_l(u)$ is the $l$th Fourier basis function, and $\xi _{\bs,t}^l$ is the coefficient for $\nu_l(u)$ at location $\bs$ and time point $t$.
 
        Define  $\bxi _{t}^l = (\xi_{\bs_1,t}^l, \xi_{\bs_2,t}^l, \dots,\xi_{\bs_N,t}^l)$ for any $l$ between 1 and $L$, and 
        assume $\bxi _{t}^l \sim N(\textbf{0}_N, \frac{1}{2}\frac{1}{m^3}\bSigma)$. To ensure curve smoothness, we set $m=1$ if $l=1$, $m=\frac{l}{2}$ if $l$ is even, and $m=\frac{l-1}{2}$ if $l$ is odd and $l\geq 3$. The derivation of $m$ and details of basis functions are deferred to Appendix \ref{app: appendix data generation}. To ensure the error functions be spatially correlated, we assume that the $N\times N$ matrix $\bSigma$ is governed by $\bSigma(\phi)_{ij}=\exp\left(-\frac{||\bs_{i}-\bs_{j}||}{\phi}\right)$ for a range parameter $\phi$. 
\paragraph{Break time:}
For the region of the $N_a=45$ true alternative locations, $\md_a:=\{\mathbf{s}_1, \ldots, \mathbf{s}_{N_a}\}$, we first generate the scaled break times ($\widetilde{k}^*_\bs: \bs \in \md_a$) from a truncated multivariate normal distribution such that $0.15 \leq \widetilde{k}^*_\bs \leq 0.85$ for any $\bs \in \md_{a}$:$$\widetilde{\boldsymbol{k}}^*=(\widetilde{k}^*_{\mathbf{s}_1}, \dots, \widetilde{k}^*_{\mathbf{s}_{N_a}})^T \sim TN( \mathbf{0.5}_{N_a}, \bSigma_a, \mathbf{0.15}_{N_a},  \mathbf{0.85}_{N_a}),$$
where 
$\mathbf{x} \sim TN(\boldsymbol{\mu}, \bSigma_a, \mathbf{b}_l, \mathbf{b}_u)$ means 
$$f(\mathbf{x}, \boldsymbol{\mu}, \bSigma_a, \mathbf{b}_l, \mathbf{b}_u)=\frac{\exp \left\{-\frac{1}{2}(\mathbf{x}-\boldsymbol{\mu})^{T} \bSigma_a^{-1}(\mathbf{x}-\boldsymbol{\mu})\right\}}{\int_{\mathbf{b}_l}^{\mathbf{b}_u} \exp \left\{-\frac{1}{2}(\mathbf{x}-\boldsymbol{\mu})^{T} \bSigma_a^{-1}(\mathbf{x}-\boldsymbol{\mu})\right\} d \mathbf{x}}.$$
Again, $\bSigma_a(\phi)_{ij}=\exp\left(-\frac{||\mathbf{s}_{i}-\mathbf{s}_{j}||}{\phi}\right)$ for $\bs_i,\ \bs_{j}\in \md_a.$
        Then the real break time $k^*_{\mathbf{s}}=[\widetilde{k}^*_{\mathbf{s}}T],\ \bs\in\md_a$, where $[a]$ denotes rounding $a$ to its nearest integer.
        We truncate the scaled break time to ensure there are a reasonable amount of data both before and after the changepoint. This also allows the signal-to-noise ratio defined later in this section to be within a normal range.  
        
\paragraph{Change functions:} We generate change functions $\delta_\mathbf{s}, \bs\in\md_a$ as follows: 
        $$\delta_\mathbf{s} = \sum_{l=1}^L \eta _{\bs}^l \nu_l,\ \bs\in\md_a,$$
        where $\nu_l$ is the $l$th Fourier basis function and $\eta _{\bs}^l$ is the coefficient for $\nu_l$ at location $\bs$.
        Define $\bet _{l} = (\eta^{l}_{\bs_1}, \ldots,\eta^{l}_{\bs_{N_a}})^T$ for any $l$ between 1 and $L$, and assume $\bet _{l} \sim N(\rho\frac{1}{m^2}\mathbf{1}_{N_a}, \frac{1}{10}\frac{1}{m^3}\bSigma_a)$, where $m$ and $\bSigma_a$ follow the definition in the error function and break time, respectively. The parameter $\rho$ measures the magnitude of the change signal. 
        
To investigate how our model performs under different spatial correlation strengths, we consider both $\phi=2$ and $5$ which corresponds to relatively weaker and stronger spatial correlation. It is also interesting to study the influence of change signal strength on our model performance. We adopt the signal-to-noise ratio (SNR) used in \cite{aue2018detecting} to measure the strength of the change signal. SNR, the ratio of the magnitude of change function to that of error functions, is defined as
\begin{equation}
\label{SNR}
    SNR = \frac{\theta (1-\theta)\|\delta\|^2}{tr(\bf{C}_\epsilon)},
\end{equation}
where $\theta$ is the scaled date of the changepoint, i.e. $k_\bs^*/T$ in our context,
 $\delta$ is the change function, $\bf{C}_\epsilon$ is the long-run covariance matrix of the error functions as defined in Equation (\ref{eq: LR_cov}), and $tr(\cdot)$ is the trace function. The estimation procedure for SNR at a single location is detailed in  \cite{aue2018detecting}. By setting $\rho=1$ and $1.5$ we obtain simulated data with mean SNR over all locations in $\md_a$ being around 0.5 and 1, which corresponds to weaker and stronger signal, respectively.  

\subsection{Results}
To evaluate the performance of the proposed method, we examine the rooted mean squared error (RMSE) of the changepoint estimate, the empirical coverage of the credible interval (CI), and the length of CI. For each setting of spatial correlation and SNR, we run $100$ simulations. Different locations, changepoints and functional data are generated independently in each simulation. For our Bayesian model, to make sure the MCMC chain has already converged, we try several sets of different initial values for all parameters and evaluate the difference between those chains with the Gelman–Rubin diagnostic \citep{gelman1992inference}. We also apply Geweke’s diagnostic \citep{geweke1991evaluating} to determine the burn‐in period. 
Through experimentation, we find that 20,000 MCMC iterations with a 15,000 burn-in period and thinning with step size 10, is sufficient to produce nearly iid samples from the posterior distribution. We compute 95\% credible intervals as the interval between the 2.5 and 97.5 percentiles of the posteriors for each parameter.

We compare our method to the recent Fully Functional (FF) method in \cite{aue2018detecting} and the method particularly designed for spatial functional data in \cite{gromenko2017detection} (hereinafter GKR). The GKR method mainly focuses on changepoint detection and does not provide confidence intervals. Thus, our comparison to \cite{gromenko2017detection} is only limited to comparing the accuracy of the estimation. 
The changepoint confidence interval based on the FF method is computed using the R package \texttt{fChange}.
Although all methods are applied to the functional data at 50 locations, the evaluation metrics are calculated only at the 45 true alternative locations.

\begin{figure}[h]
    \centering
    \includegraphics[width=\textwidth]{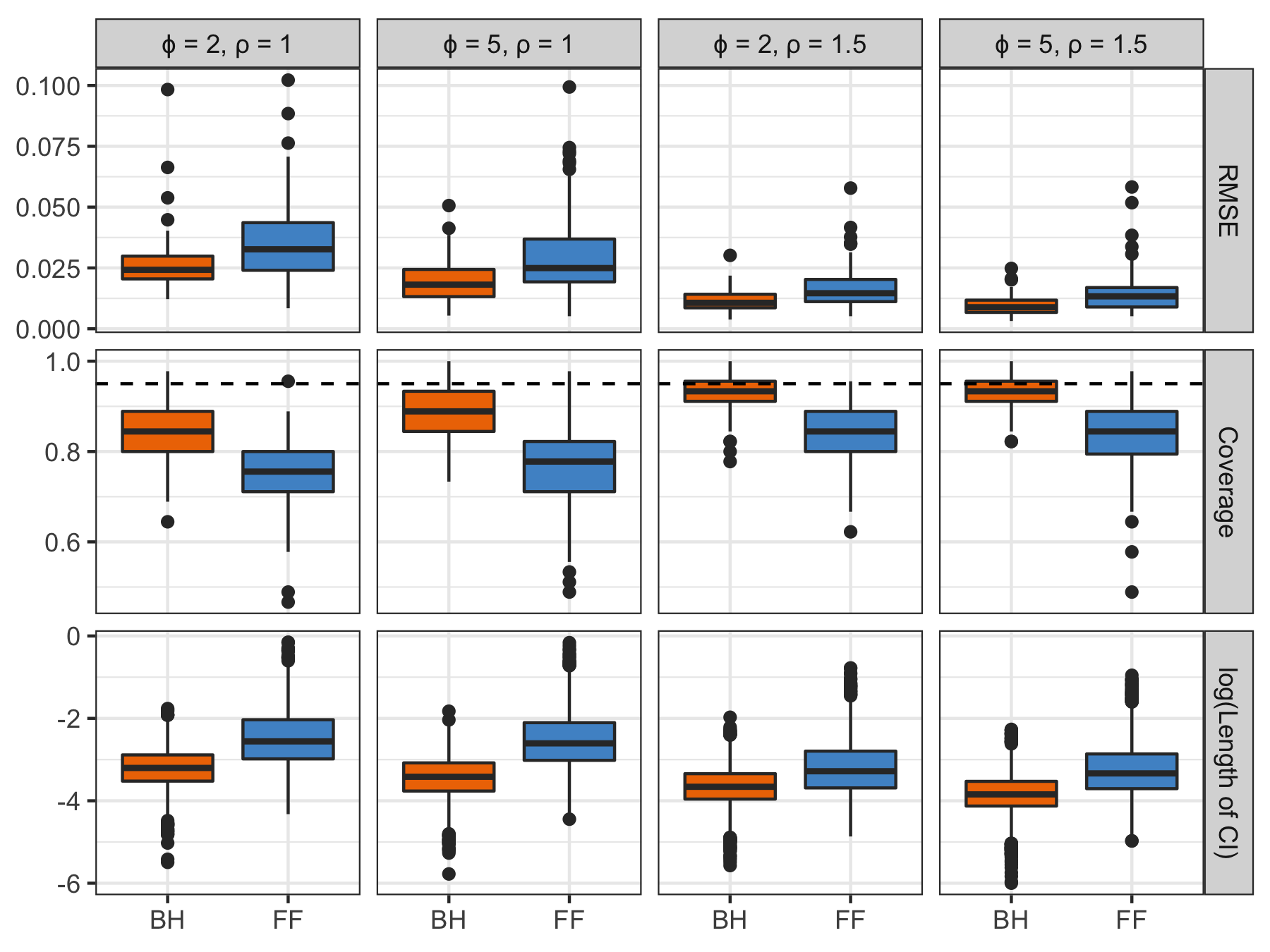}
    \caption{Boxplots of RMSE, the empirical coverage probability of 95\% credible and confidence intervals and the logarithms of interval length under four settings. The range parameter $\phi=2$ and $\phi=5$ represent weaker and stronger spatial correlation, and $\rho=1$ and $\rho=1.5$ represent weaker and stronger change signals, respectively. "BH" is our proposed Bayesian hierarchical model and "FF" refers to the fully functional method in \cite{aue2018detecting}.
    }
    \label{fig: simulation_boxplot}
\end{figure}

The RMSE of the changepoint estimates from all three methods is reported in Appendix \ref{app: simulation}.
Unsurprisingly, GKR has significantly higher error rates than the other two methods since it assumes a single changepoint whereas the data are generated with spatially varying changepoints.
We instead focus on FF and our method in Figure \ref{fig: simulation_boxplot} since they have comparable error rates. 
Across all four scenarios representing both the weaker and stronger spatial correlation and SNR, our proposed method outperforms 
FF by reducing the RMSE of the changepoint estimation. 
When the signal of change is stronger ($\rho=1.5$), both FF and our method show smaller and more stable RMSE, as expected.
When spatial correlation is higher ($\phi=5$), our method achieves far less estimation error, especially in the challenging situation with a weaker change signal ($\rho=1$). This implies that our method can use spatial correlation to improve the changepoint estimation. Curiously, the FF method experiences a slight RMSE reduction in the high correlation regime, which turned out to be an artifact of the data generation randomness. Details are reported in Appendix \ref{app: simulation}.
 
We further report the empirical coverage probability of our 95\% credible intervals 
against the 95\% confidence intervals of the FF method, and present the interval lengths of both methods in Figure \ref{fig: simulation_boxplot}. 
Narrow credible or confidence intervals, with empirical coverage close to the nominal level, indicate precise uncertainty quantification. Our credible intervals, based on weakly informative priors, are closer to the nominal level and narrower than the corresponding FF confidence intervals.
We observe that when the change signal is stronger, both FF and our method improve the uncertainty quantification compared to the lower change signal scenarios. 
Again, our method is apparently able to take advantage of the spatial correlation in changepoint estimation, reflected by shorter credible interval length while better coverage when the spatial correlation becomes stronger. 
This ability is particularly important when the change signal is weak, because in such cases, methods like FF that do not take spatial correlation into account may face challenges. 
 
\begin{figure}[t!]
\begin{subfigure}{\textwidth}
  \centering
  \includegraphics[width=\linewidth]{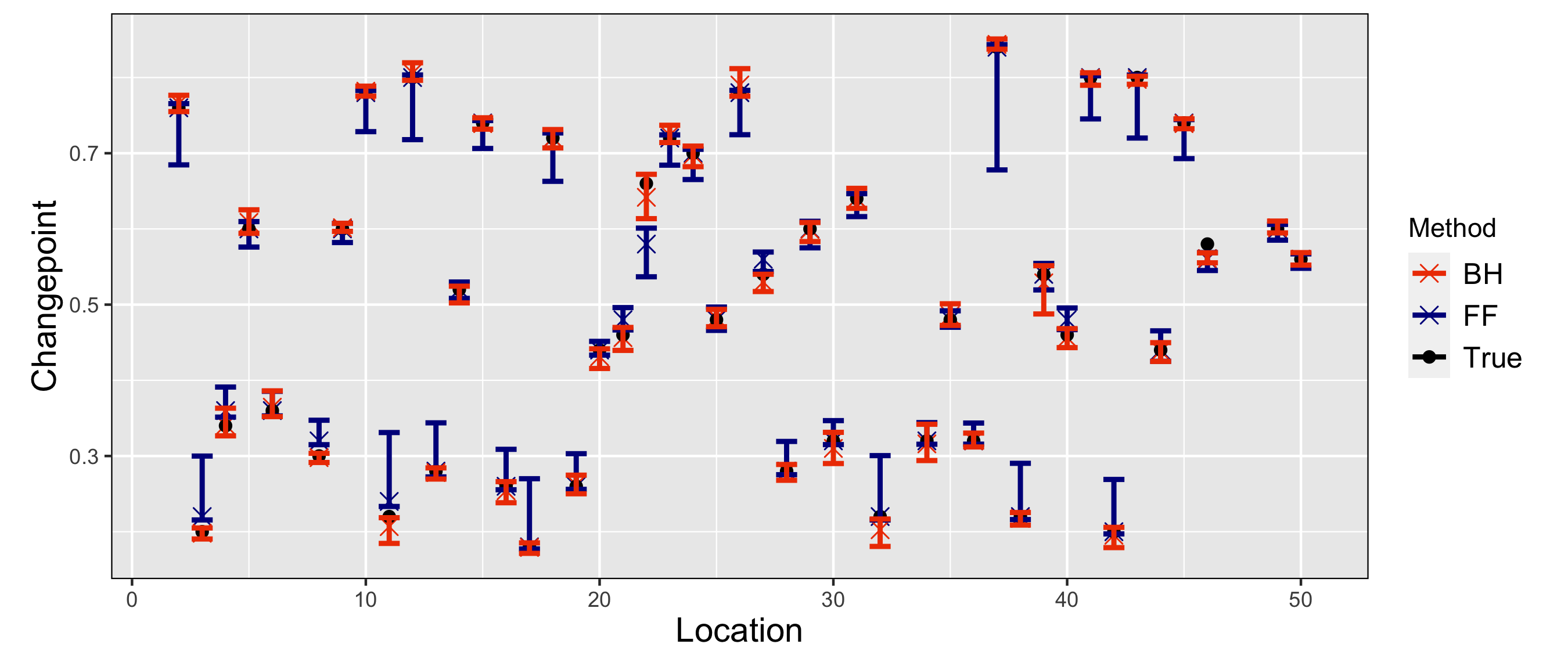}  
  \caption{}
  \label{fig: sim_CI_1}
\end{subfigure}
\begin{subfigure}{\textwidth}
  \centering
  \includegraphics[width=\linewidth]{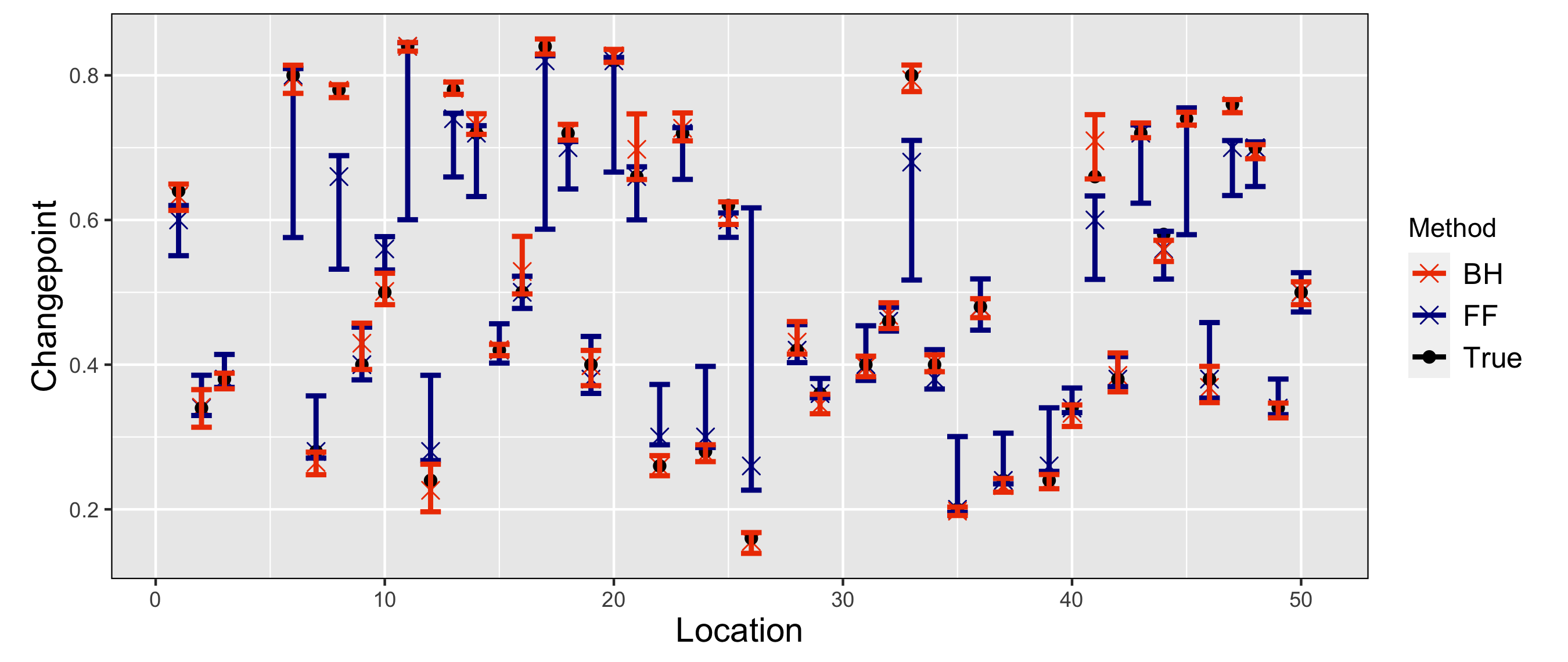}  
  \caption{}
  \label{fig: sim_CI_2}
\end{subfigure}
\caption{95\% credible 
and confidence intervals (vertical "I"), changepoint estimates (cross), and true changepoints (black dot) at alternative locations. Labeling of the procedures is the same as that in Figure \ref{fig: simulation_boxplot}. (a) A simulation from the setting $\phi = 5$ and $\rho = 1.5$. Our model has coverage 93.3\% and RMSE 0.0068, while the FF method has coverage 80\% and RMSE 0.0146. (b) A simulation from the setting $\phi = 5$ and $\rho = 1$. Our model has coverage 93.3\% and RMSE 0.0130, while the FF method has coverage 66.7\% and RMSE 0.0363.}
\label{fig: ex_CI}
\end{figure}

Figure \ref{fig: ex_CI} shows the 95\% credible 
and confidence intervals from randomly chosen simulation runs in two different settings. Figure \ref{fig: sim_CI_1} is associated with the stronger spatial correlation and the stronger change signal when both our and the FF method have the best performance among all the settings in terms of both the accuracy of the changepoint estimation and the uncertainty quantification. In this scenario, the performance of the confidence interval from the FF method is slightly worse than
the credible interval of our method and the RMSE from FF is competitive. Nevertheless, it is still seen that when the true changepoints are closer to the edges, the FF method tends to miss true values and results in longer credible intervals, while our method consistently captures all changepoints well regardless of their positions. 
Besides, for many locations, even though the estimate from the FF method is close to the true value, their confidence interval often appears too long to be informative. Figure \ref{fig: sim_CI_2} corresponds to the case with the stronger spatial correlation and weaker signal. Both methods perform satisfactorily when the true changepoint is near 0.5. However, the FF method in this scenario struggles to capture the changepoint as well as quantify the uncertainty when the real changepoint is slightly extreme toward both ends. In contrast, our method still retains its power in those situations by providing accurate estimates and informative credible intervals.  

Under the null hypothesis of no changepoint, the variability of $Y_{T,k}$ process is large in the middle and reaches its peak at 0.5. Even if the changepoint exists, the variance in the middle still tends to be higher due to the intrinsic properties of $Y_{T,k}$, though the peak may not occur at the center. Since the FF method only searches for the maximum value of the $Y_{T, k}(\bs)$ process, it could be vulnerable to the large variance often dwelling around the center of the duration.  When the real changepoint is off-center and the signal is weak, 
high variance near the center can lead to spurious maxima in the $Y_{T,k}$ process.
In contrast, our method attempts to 
identify the changepoint with a piecewise linear model, which is more robust to variance.
Furthermore, our method allows us to borrow the neighborhood information to estimate the changepoint, which is particularly helpful for challenging situations such as change signal being weak or changepoints close to the edges.   


\section{Real Data Examples}
\label{sec: real}
We demonstrate our method on two datasets 
and, again, compare
our results with the FF detector of \cite{aue2018detecting}. The first dataset is the temperature profiles introduced in Section \ref{sec: intro}, and the second 
dataset records
COVID-19 positive cases by age in Illinois during the spring of 2021.

\subsection{California Minimum Temperature}
\label{CAtemp}
As described in Section \ref{sec: intro}, we have daily minimum temperature profiles at 207 locations in California from 1971 to 2020. Due to the high degree of missingness in many sites, we only retain 28 stations that have at least 85\% complete profiles each year. Each profile is then smoothed with 21 Fourier basis functions. We apply the FF method to further subset the number of stations down to 16, each with a p-value below the 0.05 cutoff after FDR correction.

\begin{figure}[ht!]
\begin{subfigure}{.54\textwidth}
  \centering
  \includegraphics[width=\linewidth]{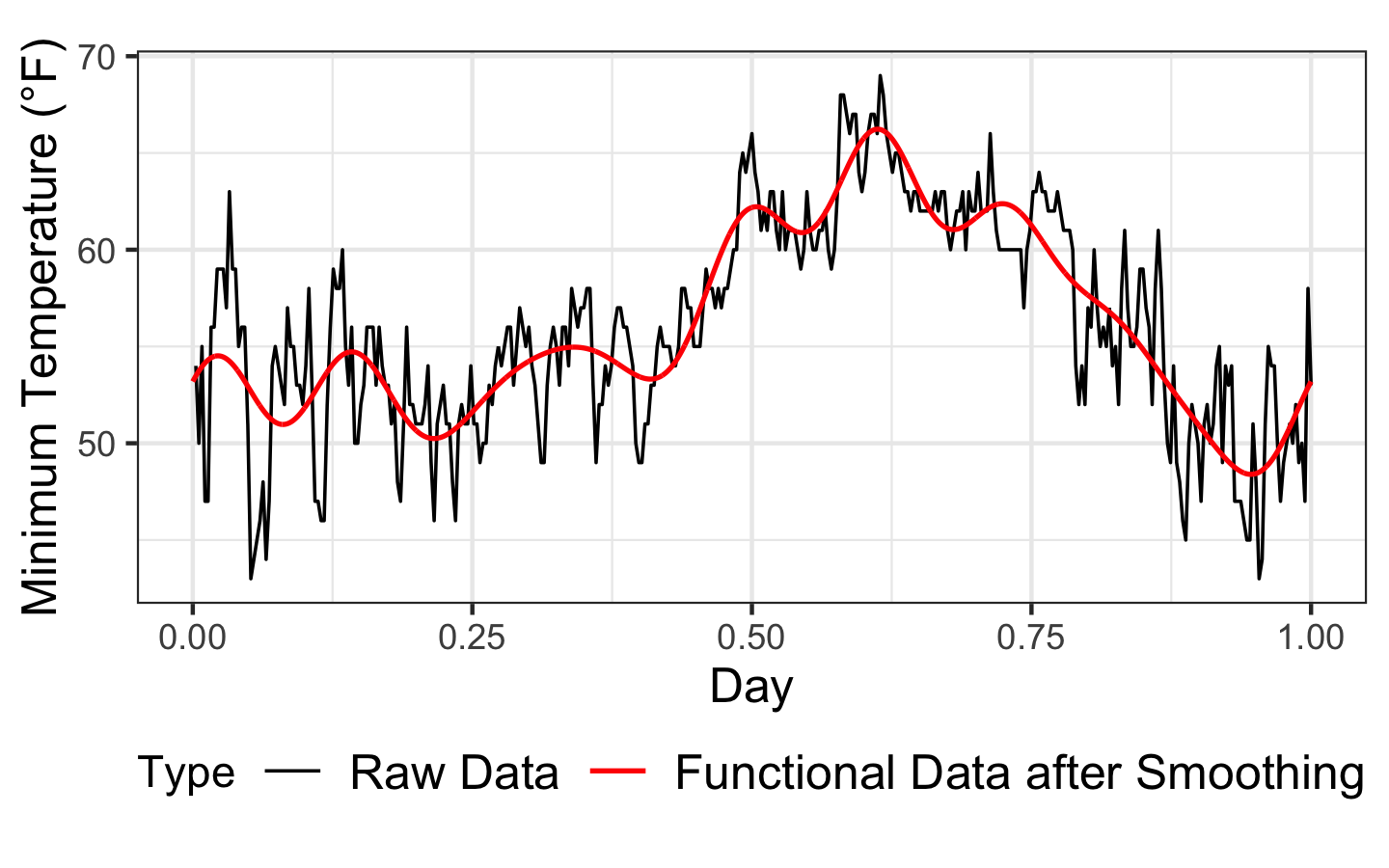}  
  \caption{}
  \label{fig: EX_temp_fd}
\end{subfigure}
\begin{subfigure}{.45\textwidth}
  \centering
  \includegraphics[width=0.95\linewidth]{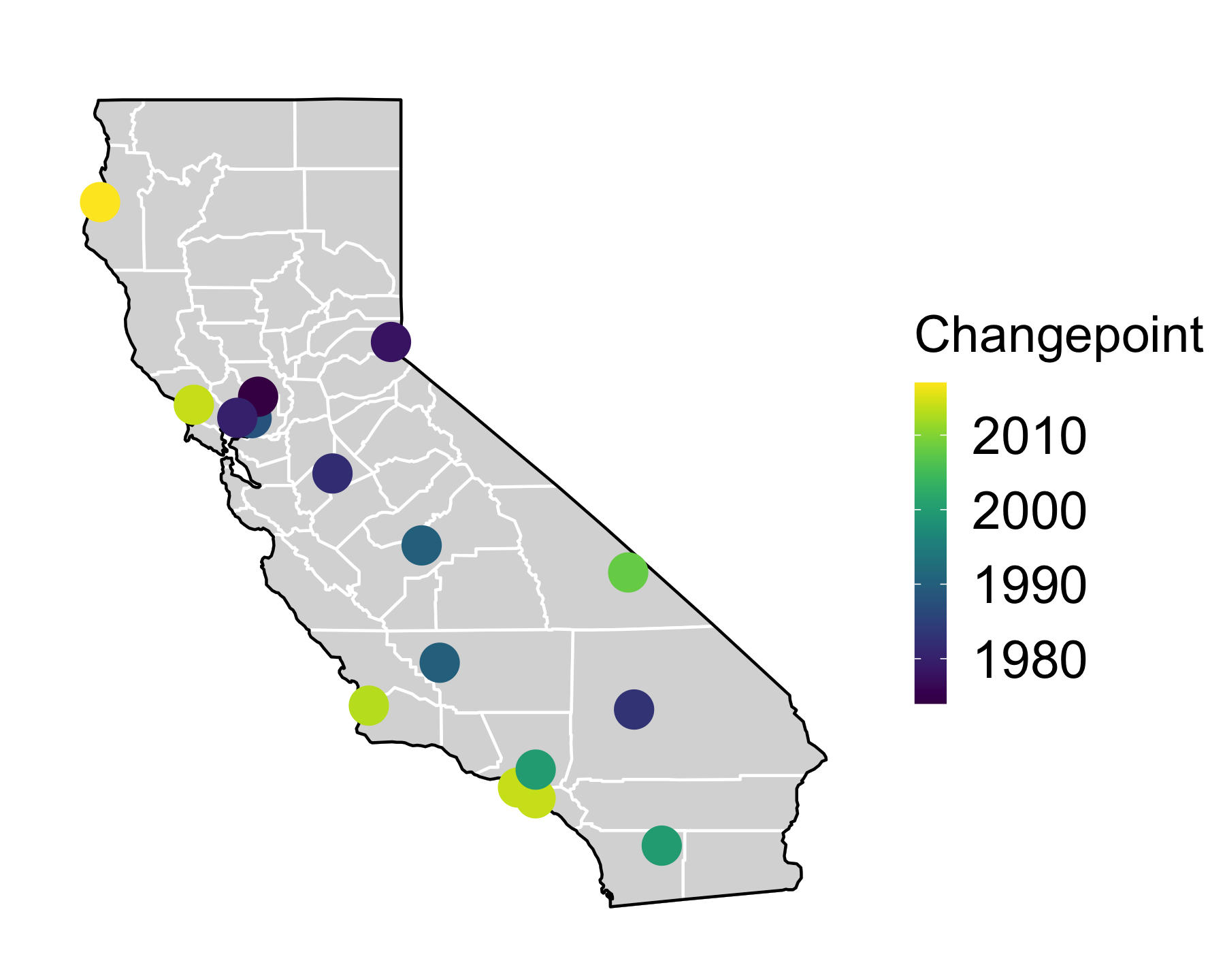}  
  \caption{}
  \label{fig: ca_BH_cp}
\end{subfigure}
\begin{subfigure}{\textwidth}
  \centering
  \includegraphics[width=\linewidth]{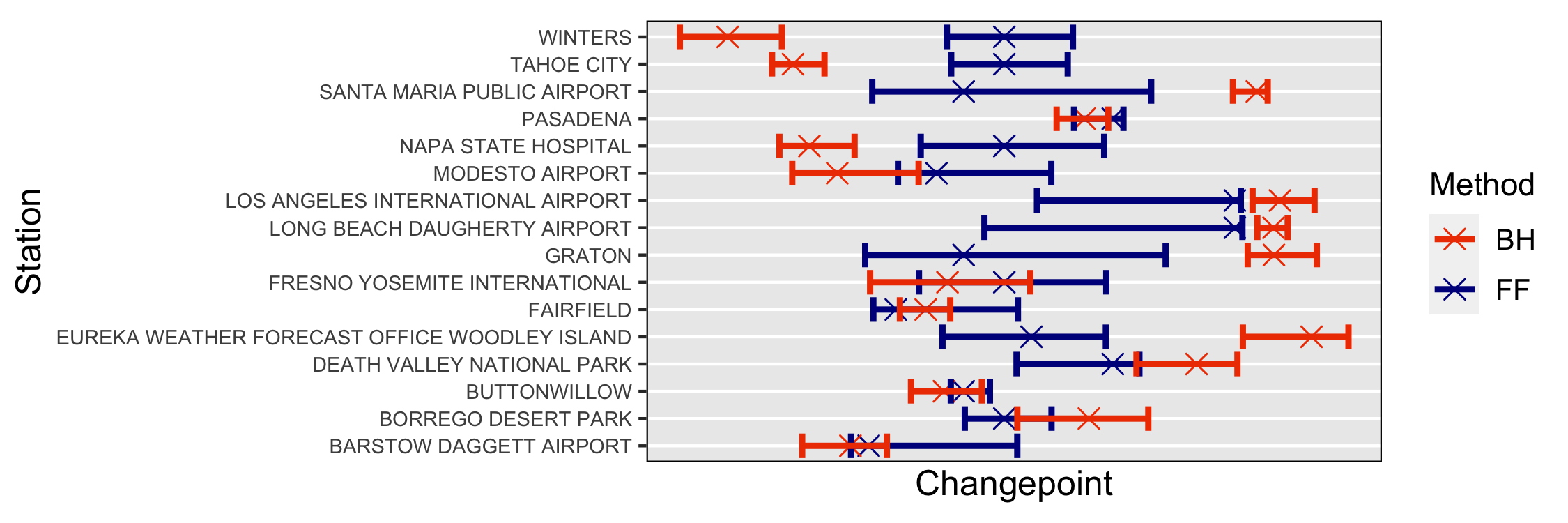}
  \caption{}
  \label{fig: ca_BH_CI}
\end{subfigure}
\caption{(a) Daily minimum temperature profile at Los Angeles International Airport in Year 1980 (black) and the smoothed curve using 21 Fourier basis functions (red). (b) Changepoint estimates from our proposed method. (c) 95\% credible intervals from our proposed method and 95\% confidence intervals from the FF method.}
\label{fig: ca_result}
\end{figure}

As an example, the daily minimum temperature profile at Los Angeles International Airport in 1980 together with the smoothed curve using the 21 Fourier basis functions are shown in Figure \ref{fig: EX_temp_fd}. When applying our method to this data, we check the MCMC convergence using the same diagnostics as discussed in the simulation, which guides us to run 30,000 iterations with a burn in of 20,000 and thinning interval of 10. The changepoint posterior estimates are presented in Figure \ref{fig: ca_BH_cp}.  We also show the credible intervals for each station in Figure \ref{fig: ca_BH_CI}, together with the FF estimates and their confidence intervals.

A comparison between Figure \ref{fig: ca_BH_cp} and Figure \ref{fig: TempCA_FF_cp}, corroborated by Figure \ref{fig: ca_BH_CI}, indicates the FF break date estimates concentrate near the middle of the interval, while our method freely finds changepoints all along the interval. The same phenomenon was observed in the simulation studies. Our estimates also preserve the spatial continuity of the naturally dependent temperature process, as evidenced by the changepoint locations in Figure \ref{fig: ca_BH_cp}.
Stations close in space tend to have changepoints close in time.
Accurate changepoint estimates and informative credible intervals can help us more profoundly understand the climate dynamics and the threat of tipping points in the climate system.

\subsection{COVID-19 Data in Illinois}
\label{sec: covid}
As we all know, the coronavirus emerged as mainly attacking the older adults, but then it is observed that the age distribution of COVID-19 cases moved toward younger ones. One interesting question in studying how COVID-19 cases evolve is to identify when the age distribution changes. To investigate this question in our state, we obtain the daily COVID-19 cases for all counties in Illinois between 01/01/2021 and 04/05/2021, 95 days in total, from the Illinois Department of Public Health (\texttt{https://www.dph.illinois.gov/covid19/data-portal}). The data reports the number of cumulative confirmed and probable positive cases in 9 age groups ($<20$, $20-29$, $30-39$, $40-49$, $50-59$, $60-69$, $70-79$, $80+$, Unknown). 
After exploratory data analysis, we eliminate the age group "Unknown" because this category only contains very few cases and the numbers are often incoherent.  

For each county, we first calculate the daily new cases for each age group and then scale them by the total number of daily new cases to approximate the age distribution. We consider the daily age distribution over time as a functional time series 
and our goal is to detect and locate any changepoints. 
We smooth the data using 7 Fourier basis functions and an example of smoothed data is shown in Appendix \ref{app: realdata}. Again, we first use the FF test and FDR control to identify the counties that show evidence of change; 28 such counties are identified. 
Champaign County has the adjusted p-value 0.102, only barely above the threshold 0.1. Since Champaign County is the 10th largest among the 102 counties in Illinois in terms of population, and it has a large young age group due to a major public university being in this county, we also include Champaign for changepoint estimation. 

We apply both FF and our method to the data over the counties that are expected to have changepoints. 
For simplicity, we use an exponential covariance function to model the dependence between county level parameters, and use the county geographical center to calculate distance, though conditional or simultaneous autoregressive models are usually more typical for areal aggregated data. We do not expect the results will be sensitive to the choice of the covariance model due to the scatter of the 29 counties. Using the same convergence diagnostic as for the previous temperature dataset, we run MCMC for 50,000 iterations and take the first 40,000 as the burn-in, then we thin the rest using the stepsize 10 to obtain the posterior samples. 

\begin{figure}[h]
\begin{subfigure}{.5\textwidth}
  \centering
  \includegraphics[width=\linewidth]{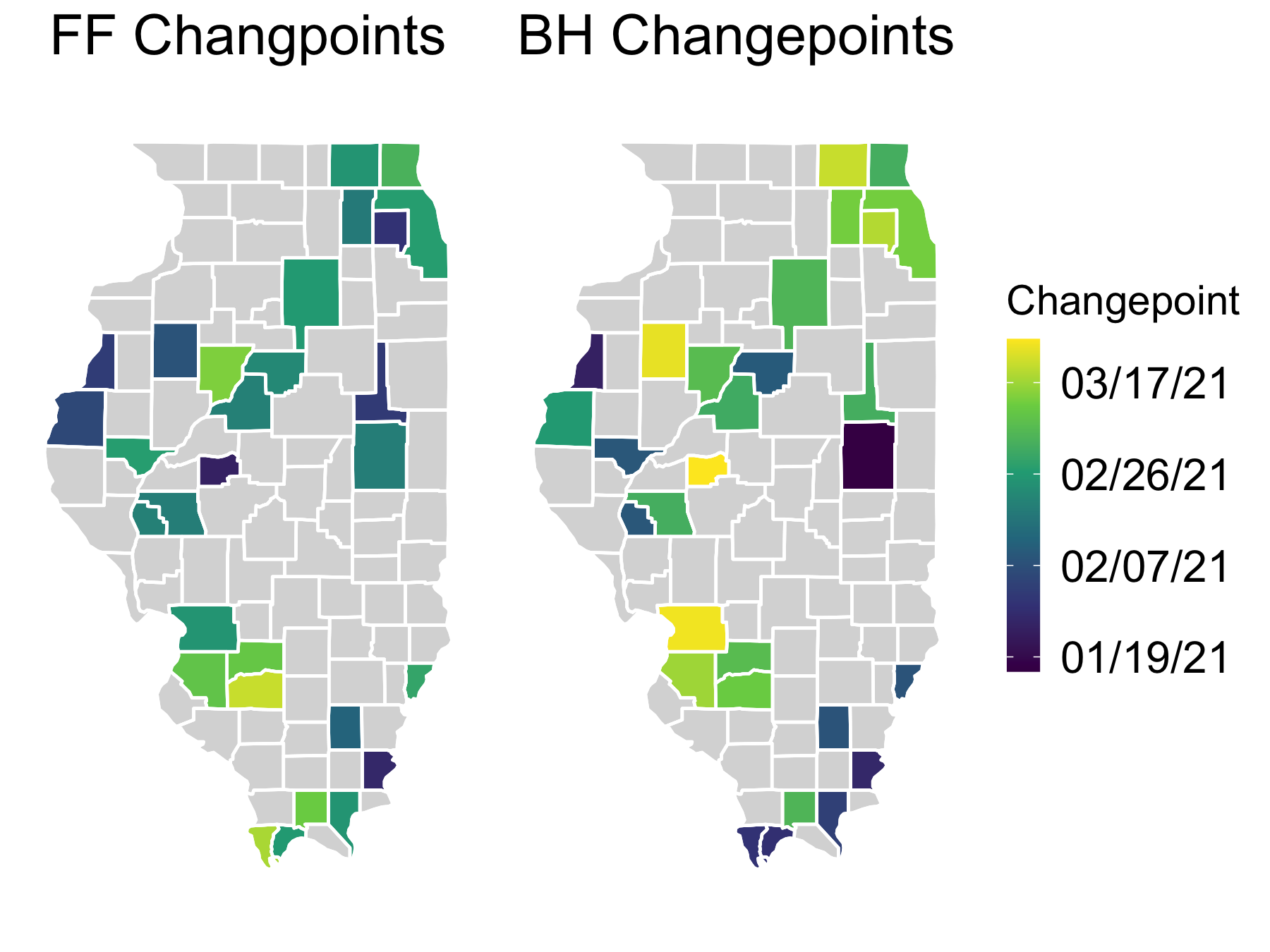}  
  \caption{}
  \label{fig: covid_cp}
\end{subfigure}
\begin{subfigure}{.5\textwidth}
  \includegraphics[width=\linewidth]{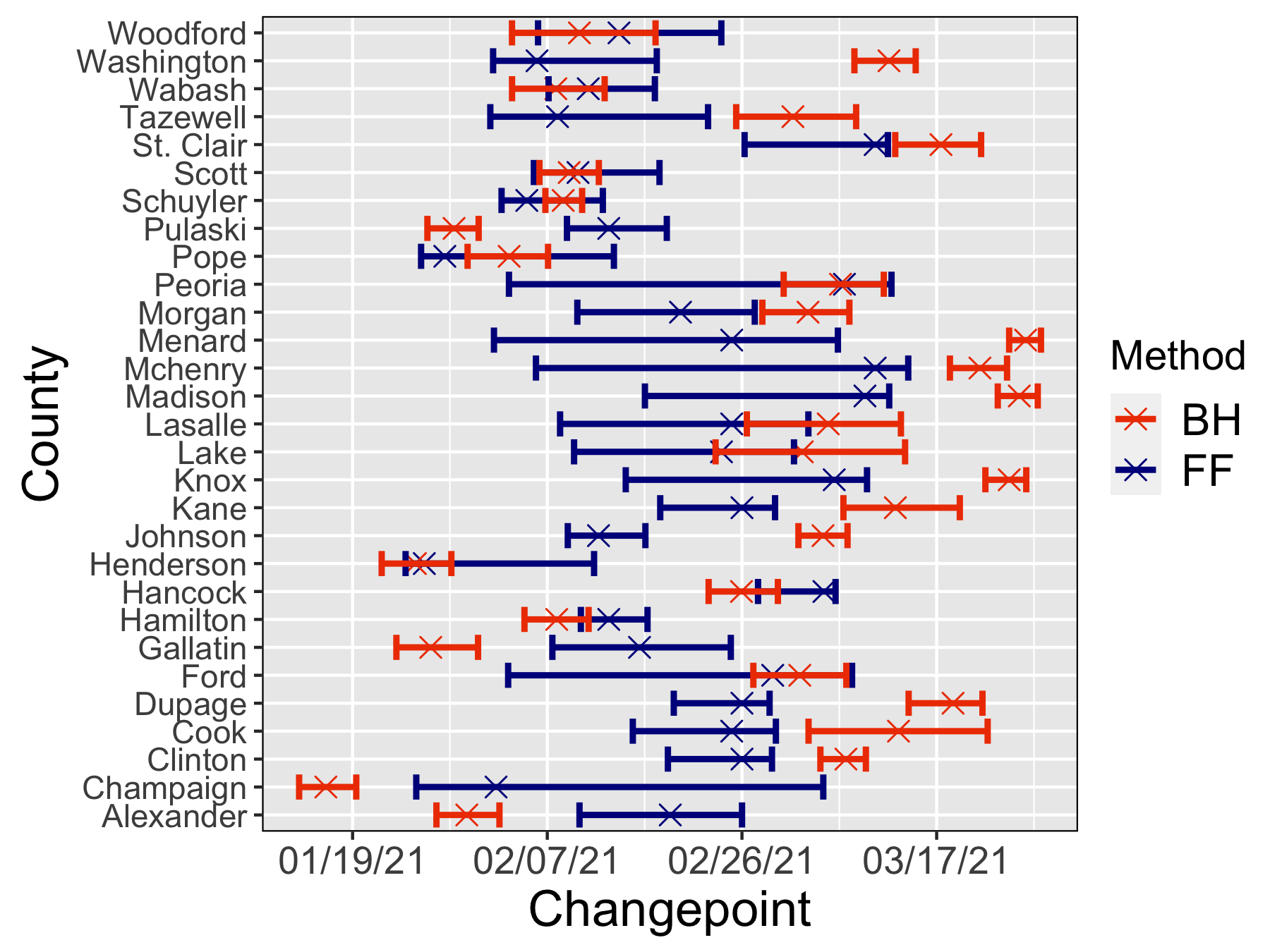}  
  \caption{}
  \label{fig: covid_CI}
\end{subfigure}
\caption{(a) Changepoint estimates from the FF method (left) and our proposed BH method (right) over the 29 counties. (b) Credible intervals from our method and confidence intervals from the FF method. 
}
\label{fig: covid_result}
\end{figure}

The changepoint estimates from both methods are illustrated in Figure \ref{fig: covid_cp}, and the 95\% credible intervals and confidence intervals are shown in Figure \ref{fig: covid_CI}. Again, 
our method is able to estimate changepoints close to the boundaries
, and our credible intervals are shorter than the FF confidence intervals. 

\begin{figure}[H]
\begin{subfigure}{.495\textwidth}
  \centering
  \includegraphics[width=\linewidth]{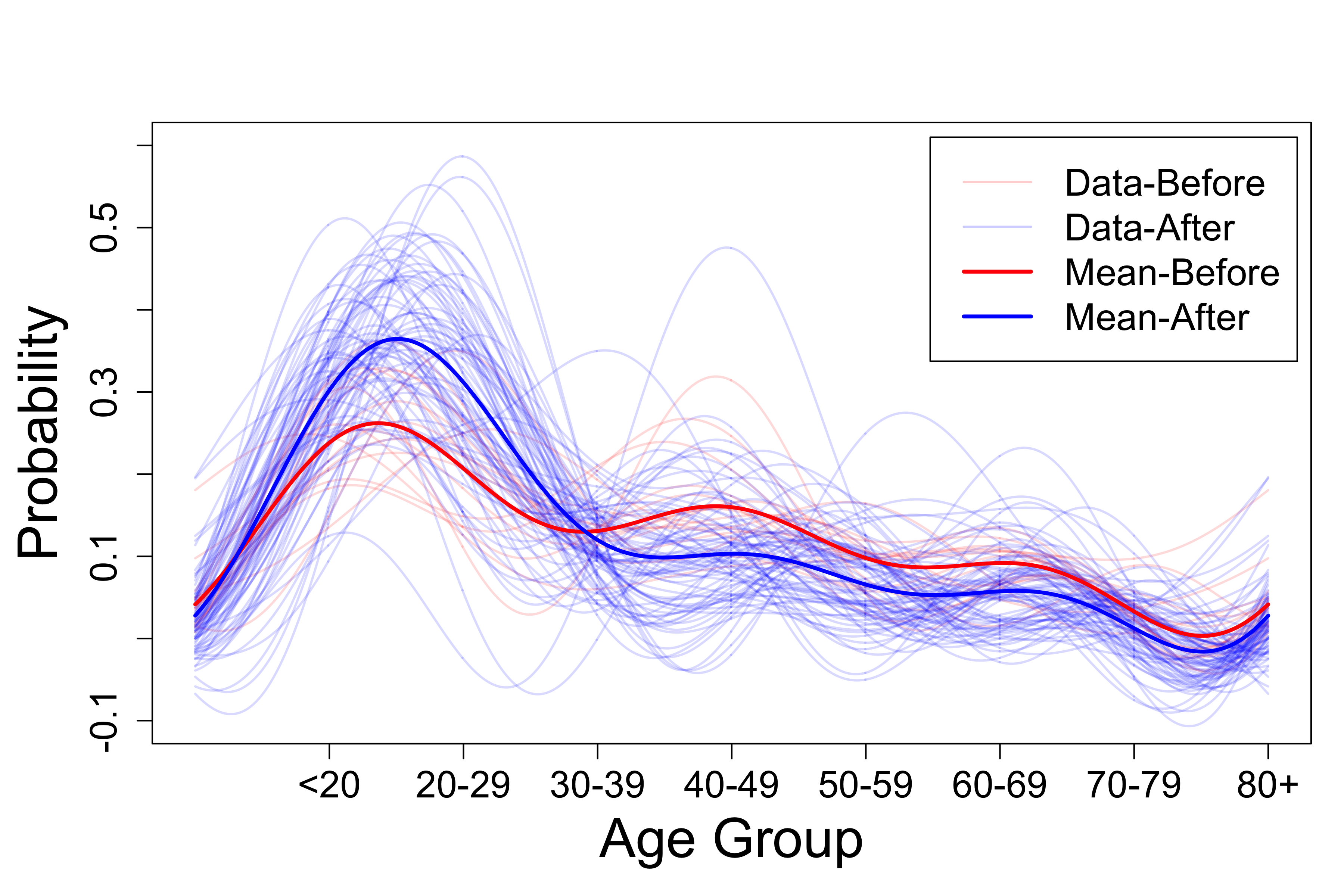}  
  \caption{}
  \label{fig: covid_ex_champaign}
\end{subfigure}
\begin{subfigure}{.495\textwidth}
  \centering
  \includegraphics[width=\linewidth]{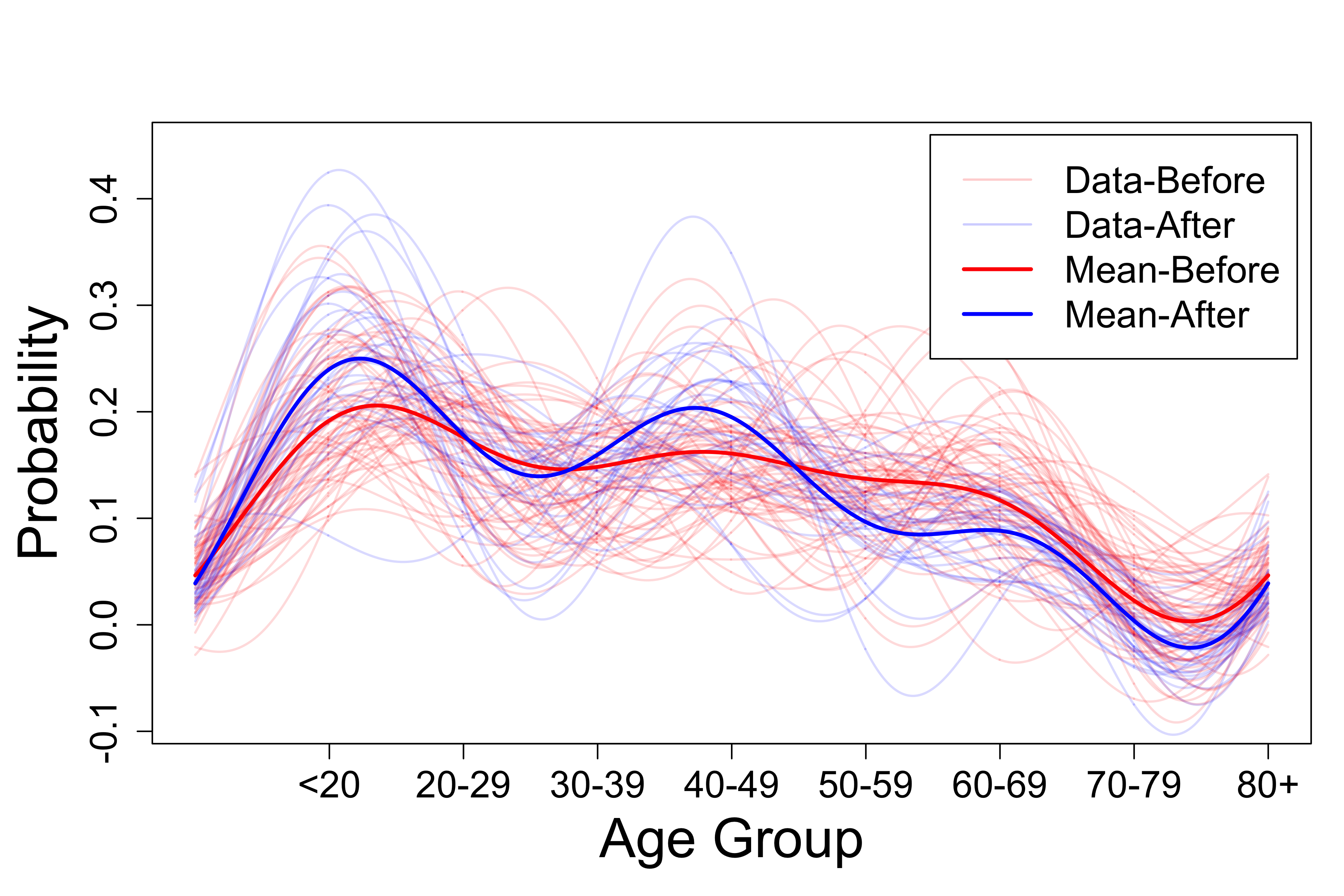}  
  \caption{}
  \label{fig: covid_ex_peoria}
\end{subfigure}
\caption{ Functional time series of COVID-19 age distribution in (a) Champaign County and (b) Peoria County. The light red and blue curves represent the functional data before and after the changepoint, respectively. The solid red and blue curves are the respective mean of the light red and blue curves.}
\label{fig: covid_ex_f_BH}
\end{figure}

To illustrate how the age distribution changes, we further plot the functional time series and their mean functions colored in two groups, whether before or after the changepoint, using Champaign and Peoria as two examples. In Figure \ref{fig: covid_ex_f_BH}, we can see in both counties, the ratio of younger people getting the coronavirus increases and that of the elder drops.
For Champaign County, where the University of Illinois Urbana-Champaign is located, a changepoint is detected on Jan. 16th, 2021 by our method. According to the school calendar, University residence halls were open for the spring semester on Jan. 17th. So it was approximately the time when students in $<20$ and $20-29$ age groups began to gather at the university. This could be one factor for cases shifting to the younger-age groups for this county.

\section{Discussion}
\label{sec: discussion}
We developed a Bayesian hierarchical model for estimating a single mean changepoint for spatially indexed functional time series. Our method allows each location to have its own changepoint but also respects the fact that the changepoints and change functions tend to be similar when the locations are close.
Simulations show that our model provides more accurate changepoint estimates and shorter but more informative credible intervals than the FF estimates and their confidence intervals.  
In particular, our method outperforms the FF method in estimating the early or late changepoints. We demonstrated our proposed method on the daily minimum temperature in California and the COVID-19 cases over age groups in Illinois.

Our method is established based on the properties of
the $Y_{T,k}$ process, a function of the CUSUM statistic that is widely employed for changepoint detection and estimation. Instead of searching for the maximum value of the $Y_{T,k}$ process as in the FF method, we proposed to use a two-piece piecewise linear model to capture the peak of the $Y_{T,k}$ process. We carefully built the variance structure of the $Y_{T,k}$ process into the piecewise linear model so that the uncertainty of the linear model fitting and thus the changepoint estimates are appropriately quantified. 
By jointly fitting the spatially correlated piecewise linear models across all locations through a Bayesian hierarchical model, we took the inherent spatial correlation into account in our changepoint estimation.

Our method differs significantly from the existing methods in two aspects. 
First, we utilize spatial correlation to synthesize information over the whole spatial domain instead of focusing on a single location, e.g., \cite{aue2018detecting}. 
Second, we allow spatially varying changepoints for different locations, 
instead of assuming a single shared changepoint across all locations (\citealp{gromenko2017detection}).
Our model essentially combines the strengths of Aue et al. and Greomenko et al. to achieve highly accurate and flexible changepoint estimation in space. We also show that our method produces precise, informative, and intuitive credible intervals of the changepoint.

Finally, our method currently only focuses on the changepoint estimation, after the rejection region of changepoint detection has been identified. In future work, we would like to develop a more compact approach by incorporating detection and estimation in a single model to remove dependence on auxiliary methods for detection. 


\bibliography{references}

\newpage
\appendix
\section{Long-run covariance kernel}
\label{app: eigen of kernel}
We first define the long-run covariance kernel, eigenvalues and eigenfunctions related to the error functions in Equation \eqref{equ: fdmodel}. Since the error properties are assumed homogeneous across all locations, we drop the spatial index and represent the error functions at one spatial location as $\varepsilon_t, t \in \mathds{Z}$ in the following. Under Assumption \ref{assumption}, the limiting performance of $Y_{T,k}(\bs)$ at location $\bs$ depends on the long-run covariance kernel of the error terms $\varepsilon_t: t \in \mathds{Z}$. The kernel is defined as
\begin{equation}
\label{eq: LR_cov}
    C_\varepsilon\left(u, u'\right) = \sum_{l=-\infty}^{\infty}\operatorname{cov}\left\{\varepsilon_0(u), \varepsilon_l(u')\right\},
\end{equation}
which was first considered by \cite{hormann2010weakly} with its estimator and convergence further studied.
A positive definite and symmetric Hilbert-Schmidt integral operator $c_{\varepsilon}$ on $L^{2}[0,1]$ can be defined using $C_\varepsilon$. More formally,
\begin{equation}
    c_{\varepsilon}(g)(u)=\int C_{\varepsilon}(u, u') g(u') \mathrm{d} u',
\end{equation}
where $g \in L^2([0,1])$. This further defines a non-increasing sequence of non-negative eigenvalues $\lambda_{l}: l \in \mathds{N}$ and the corresponding orthonormal eigenfunctions $\psi_{l}: l \in \mathds{N}$, which satisfy
\begin{equation}
\label{eq: eigen of kernel}
    c_{\varepsilon}\left(\psi_{l}\right)(u)=\lambda_{l} \psi_{l}(u), \quad l \in \mathds{N}.
\end{equation}
The eigenvalues and eigenfunctions of $c_{\varepsilon}$ determine the asymptotic mean and variance of the $Y_{T,k}(\bs)$ time series. The estimations of eigenvalues and eigenfunctions are provided by \texttt{fChange} package in R. We use the optimal bandwidth selector provided by this package to complete the estimation.

\section{Brownian Bridge Properties} 
Let $W(t)$ denote a standard Wiener process (or Brownian motion), i.e. $W(t)$ is a stochastic process such that for $t \geq 0$, the increments $W(t) - W(0)$ are stationary, independent, and normally distributed with $E\{W(t)\} = 0$ and $\operatorname{var}\{W(t)\} = t$. We can further define a Brownian bridge on $[0, T]$ as the process $B(t)=W(t)-\frac{t}{T}W(T)$ for $t\in[0,T]$.

\begin{lem}
\label{lemma:BB}
 If $B(t)$ is a Brownian bridge for $t\in[0,1]$, then it has the following properties:
 $$E\{B^2(t)\} =t\left(1-t\right),$$
 $$\operatorname{var}\{B^2(t)\}=2t^2(1-t)^2,$$
 $$\operatorname{cov}\{B^2(t), B(t)\}=0.$$
\end{lem}
To simplify the notation, we write $B(t)$ as $B_t$ and $W(t)$ as $W_t$. 
According to the definition and properties of Brownian bridge, it can be seen that $E(B_t)=0$,
$\operatorname{var}\left(B_t\right)=\frac{t(T-t)}{T}$ and $\operatorname{cov}(W_s, W_t)=E(W_s W_t)=s,\ s\leq t$.
Furthermore, we can get the following equations,
\begin{equation}
\begin{aligned}
 E(B^2_t) &= E\left\{\left(W_t-\frac{t}{T}W_T\right)^2\right\}= E\left(W^2_t+\frac{t^2}{T^2}W^2_T-\frac{2t}{T}W_t W_T\right)\\
 &=t+\frac{t^2}{T^2}T-\frac{2t}{T}t=t+\frac{t^2}{T}-\frac{2t^2}{T}=t-\frac{t^2}{T},
\end{aligned}
\end{equation}

\begin{equation}
    \begin{aligned}
    \operatorname{var}(B_t^2) &= \operatorname{var}\left\{\left(W_t-\frac{t}{T}W_T\right)^2\right\}=\operatorname{var}\left(W_t^2+\frac{t^2}{T^2}W_T^2-\frac{2t}{T}W_t W_T\right)\\
    &= \operatorname{var}(W_t^2)+\frac{t^4}{T^4}\operatorname{var}(W_T^2)+\frac{4t^2}{T^2}\operatorname{var}(W_t W_T)+\frac{2t^2}{T^2}\operatorname{cov}(W_t^2, W_T^2)\\
    & -\frac{4t}{T}\operatorname{cov}(W_t^2, W_t W_T)-\frac{4t^3}{T^3}\operatorname{cov}(W_T^2,W_t W_T).
    \end{aligned}
    \label{equ: varB_t}
\end{equation}

Since $\frac{W_t}{\surd{t}}\sim N(0,1)$, $E\left( \frac{W_t^2}{t} \right)=1$ and $E\left( \frac{W_t^4}{t^2} \right)=3$, i.e.$
    E\left(W_t^2 \right)=t$, $E\left(W_t^4 \right)=3t
    ^2$, we derive the following equations,
$$\operatorname{var}\left(W_t^2 \right)=E(W_t^4)-E^2(W_t^2)=2t^2,$$
\begin{equation}\nonumber
\begin{aligned}
    \operatorname{var}(W_t W_T)&=E(W_t^2W_T^2)-E^2(W_t W_T)
    =E\left[W_t^2\{W_t+(W_T-W_t)\}^2\right]-E^2(W_t W_T)\\
    &=E\{W_t^4+2W_t^3(W_T-W_t)+W_t^2(W_T-W_t)^2\}-E^2(W_t W_T)\\
    &=3t^2+t(T-t)-t^2
    =tT+t^2,
\end{aligned}
\end{equation}


\begin{equation}\nonumber
    \operatorname{cov}(W_t^2, W_T^2)=E(W_t^2W_T^2)-E(W_t^2)E(W_T^2)=tT+2t^2-tT=2t^2,
\end{equation}

\begin{equation}\nonumber
    \begin{aligned}
    \operatorname{cov}(W_t^2, W_t W_T)&=E(W_t^3 W_T)-E(W_t^2)E(W_t W_T)\\
    &=E[W_t^3\{W_t+(W_T-W_t)\}]-E(W_t^2)E(W_t W_T)\\
    &=E(W_t^4)-E(W_t^2)E(W_t W_T)
    = 3t^2-t^2
    = 2t^2,
    \end{aligned}
\end{equation}

\begin{equation}\nonumber
    \begin{aligned}
    &\operatorname{cov}(W_T^2, W_t W_T)\\
    =&E(W_T^3W_t)-E(W_T^2)E(W_t W_T)\\
    =&E[W_t\{W_t+(W_T-W_t)\}^3]-E(W_T^2)E(W_t W_T)\\
    =&E[W_t\{W_t^3+3W_t^2(W_T-W_t)+3W_t (W_{T}-W_{t})^2+(W_{T}-W_{t})^3\}]-E(W_T^2)E(W_t W_T)\\
    =&E(W_t^4)+3E\{W_t^2(W_T-W_t)^2\}-E(W_T^2)E(W_t W_T)\\
    =&3t^2+3t(T-t)-Tt\\
    =&2tT.
    \end{aligned}
\end{equation}

After plugging the result of each item into Equation (\ref{equ: varB_t}), we can have
\begin{equation}\nonumber
    \operatorname{var}(B_t^2)= \frac{2t^2}{T^2}(T-t)^2.
\end{equation}

In this paper, we only consider the Brownian bridge on the unit interval $[0,1]$, i.e. $T=1$. Therefore, the result is further simplified as 
 $$E(B^2_t) =t\left(1-t\right),$$
\begin{equation}\nonumber
    \operatorname{var}(B_t^2)=2t^4-4t^3+2t^2=2t^2(1-t)^2,
\end{equation}
\begin{equation}
\begin{aligned}
\operatorname{cov}\left(B_t^2,B_t\right)&=E(B_t^3)-E(B_t^2)E(B_t)=E(B_t^3)=E\left\{\left(W_t-\frac{t}{T}W_T\right)^3\right\}\\
&=E\left(W_t^3-\frac{t^3}{T^3}W_T^3-\frac{3t}{T}W_t^2W_T+\frac{3t^2}{T^2}W_t W_T^2\right)\\
&=-\frac{3t}{T}E(W_t^2W_T)+\frac{3t^2}{T^2}E(W_tW_T^2),
\end{aligned}
\label{equ: Btcov}
\end{equation}
\begin{equation}\nonumber
    E(W_t^2W_T)=E[W_t^2\{W_t+(W_T-W_t)\}]=0,
\end{equation}
\begin{equation}\nonumber
    E(W_t W_T^2)=E[W_t\{W_t^2+2W_t (W_T-W_t)+(W_T-W_t)^2\}]=0.
\end{equation}
Plugging the results of the above two equations into Equation (\ref{equ: Btcov}), we can get $\operatorname{cov}\left(B_t^2,B_t\right)=0$.

\section{Proof of Lemma \ref{thm:m&v_H0} and Proposition \ref{thm:m&v_HA}}
\label{app: proof}
To simplify notation, we drop the location index from Equation \ref{equ: fdmodel} and assume the observations, at location $\bs$, follow
$$X_{t}(u) = \mu(u) + \delta(u)\mathds{1}(t>k^* )+\epsilon_{t}(u),\ t=1,\dots, T.$$
The hypothesis about changepoint detection is
$$
H_{0}: \delta(u)=0 \quad \text { versus } \quad H_{A}: \delta(u) \neq 0,
$$
and the definition of CUSUM statistic is

$$S_{T, k}(u)= \frac{1}{\surd{T}}\left\{\sum_{t=1}^k X_{t}(u)-\frac{k}{T}\sum_{t=1}^{T} X_{t}(u)\right\}, \ t=0,\dots, T.$$
For spatial locations with no changepoint, i.e. null locations, we denote the CUSUM statistic by $S_{T, k}^0(u)$. For locations with a changepoint, i.e. alternative locations, we denote the CUSUM statistic as $S_{T, k}^A(u)$.

Recall Theorem 1.2 of \cite{jirak2013weak} which states that, for
 $$S_{T}(q, u)=\frac{1}{\surd{T}} \sum_{t=1}^{\lfloor T q\rfloor} \varepsilon_{t}(u),$$
under Assumption \ref{assumption}, there exists a sequence of Gaussian processes, 

$\left(\Gamma_{T}(q, u): T \in \mathds{N}, q, u \in[0,1]\right)$,
such that $E\left\{\Gamma_{T}(q, u)\right\}=0$, 
$$E\left\{\Gamma_{T}(q, u) \Gamma_{T}\left(q^{\prime}, u^{\prime}\right)\right\}=\min (q, q^{\prime}) C_{\varepsilon}\left(u, u^{\prime}\right),$$ and
$$
\sup _{0 \leq q \leq 1} \int\left\{S_{T}(q, u)-\Gamma_{T}(q, u)\right\}^{2} d u=o_{p}(1).
$$

From this theorem, we immediately have that
$$\left\|S_T(q,u)-\Gamma_T(q,u)\right\|^2 =o_{p}(1),\ for\ all\ q\in[0,1].$$
Then we can have,
$$\left\|S_T(q,u)-\Gamma_T(q,u)\right\| =o_{p}(1),\ for\ all\ q\in[0,1],$$
and
$$\left|\left\|S_T(q,u)\right\|-\left\|\Gamma_T(q,u)\right\|\right| =o_{p}(1),\ for\ all\ q\in[0,1].$$

\subsection{Under $H_0$}
At a null location we have
$E(X_1)=\dots=E(X_T)=\mu$, and
\begin{equation}\nonumber
\begin{aligned}
S_{T, k}^0 &= \frac{1}{\surd{T}}\left(\sum_{t=1}^k X_{t}-\frac{k}{T}\sum_{t=1}^{T} X_{t}\right)\\
&= \frac{1}{\surd{T}}\left\{\sum_{t=1}^k(\mu+\epsilon_t)-\frac{k}{T}\sum_{t=1}^{T} (\mu+\epsilon_{t})\right\}\\
&= \frac{1}{\surd{T}}\left(\sum_{t=1}^k\epsilon_t-\frac{k}{T}\sum_{t=1}^{T} \epsilon_{t}\right).
\end{aligned}
\end{equation}
It can be easily seen that $S_{T,k}^0(u)=S_T(\frac{k}{T},u)-\frac{k}{T}S_T(1,u),\  k=0,\dots, T$.
In another way, we write $q=\frac{k}{T}$, and then $S_{T,k}^0(u)=S_{T,Tq}^0(u)$. We further define $$S_{T}^0(q,u)=S_T(q,u)-q S_T(1,u),\  q=0,\frac{1}{T},\dots, 1.$$
In this case, $S_{T,Tq}^0(u)= S_{T}^0(q,u)$.

Note that $S_{T}(\frac{k}{T}, u)=T^{-1/2} \sum_{t=1}^{k} \varepsilon_{t}(u)$ and $S_{T}(1, u)=T^{-1/2} \sum_{t=1}^{T} \varepsilon_{t}(u)$.
Corresponding to $S_{T,k}^0(u)$, we define $$\Gamma_T^0(q,u)=\Gamma_T(q,u)-q\Gamma_T(1,u),\ q\in [0,1].$$

The goal here is $\left\|S_T^0(q,u)\right\|^2$, and note that
\begin{equation}
    \begin{aligned}
     \left\|S_T^0(q,u)-\Gamma^0_T(q,u)\right\|&=\left\|\{S_T(q,u)-q S_T(1,u)\}-\{\Gamma_T(q,u)-q\Gamma_T(1,u)\}\right\|\\
     &=\left\|\{S_T(q,u)-\Gamma_T(q,u)\}-q\{S_T(1,u)-\Gamma_T(1,u)\}\right\|\\
     &\leq \left\|S_T(q,u)-\Gamma_T(q,u)\right\|+q\left\|S_T(1,u)-\Gamma_T(1,u)\right\|=o_{p}(1).
    \end{aligned}
    \label{equ: S0&Gamma0}
\end{equation}
Since $\left|\left\|S_T^0(q,u)\right\|-\left\|\Gamma^0_T(q,u)\right\|\right|\leq \left\|S_T^0(q,u)-\Gamma^0_T(q,u)\right\|$, $\left|\left\|S_T^0(q,u)\right\|-\left\|\Gamma^0_T(q,u)\right\|\right|=o_{p}(1)$. Or in another way,

\begin{equation}
    \left\|S_T^0(q,\cdot)\right\|=\left\|\Gamma^0_T(q,\cdot)\right\|+o_{p}(1).
    \label{equ:S&Gamma}
\end{equation}
Following Theorem 1 of \cite{aue2018detecting} and the proof in their supplement materials, using the definition of $\Gamma_{T}(q, u)$, calculations can be done to show that $E\left\{\Gamma_{T}^{0}(q, u) \Gamma_{T}^{0}\left(q', u^{\prime}\right)\right\}=\left\{\min (q, q^{\prime})-q q^{\prime}\right\} C_{\varepsilon}\left(u, u^{\prime}\right)$.
Hence, for all $T$, the Gaussian process $\Gamma_{T}^{0}(q, u)$ has the same distribution as
$$
\sum_{\ell=1}^{\infty} \lambda_{\ell}^{1/2} B_{\ell}(q) \phi_{\ell}(u),
$$
where $\lambda_{\ell}$ and $\phi_{\ell}$ are defined as in Appendix \ref{app: eigen of kernel}.
And $\left(B_{\ell}: \ell \in \mathds{N}\right)$ are independent and identically distributed standard Brownian bridges defined on
$[0,1]$. Following the supplement of \cite{aue2018detecting}, it is obvious that, for all $T$,
\begin{equation}\nonumber
\left\|\Gamma_{T}^{0}(q, \cdot)\right\|\stackrel{\mathcal{D}}{=}   \left\| \sum_{\ell=1}^{\infty} \lambda_{\ell}^{1/2} B_{\ell}(q) \phi_{\ell}(u)\right\|
\stackrel{\mathcal{D}}{=} \left\{\sum_{\ell=1}^{\infty} \lambda_{\ell} B_{\ell}^{2}(q)\right\}^{1/2},
\label{equ:Gamma&BB}
\end{equation}
which, in light of Equation (\ref{equ:S&Gamma}), Slutsky's theorem and continuous mapping, implies $\left\|S_T^0(q,\cdot)\right\|^2  \overset{\mathcal{D}}{\to}  \sum_{\ell=1}^{\infty} \lambda_{\ell} B_{\ell}^{2}(q),\ (T \to \infty)$, i.e. for any location \bs,
\begin{equation}
Y_{T, k}(\bs) \overset{\mathcal{D}}{\to}  \sum_{\ell=1}^{\infty} \lambda_{\ell} B_{\ell}^{2}(q)\ \ \ \ (T \to \infty).
\end{equation}
We further explore the mean and variance of its asymptotic distribution $\sum_{\ell=1}^{\infty} \lambda_{\ell} B_{\ell}^{2}(q)$, where $q \in [0,1]$.
Recall Lemma \ref{lemma:BB}, and the fact that $B_l$ are independent, we have that
\begin{equation}
\label{equ: null_mean}
    E\left\{\sum_{\ell=1}^{\infty} \lambda_{\ell} B_{\ell}^{2}(q)\right\} = \sum_{\ell=1}^{\infty} \lambda_{\ell}q(1-q),
\end{equation}
\begin{equation}
\label{equ: null_var}
    \operatorname{var}\left\{\sum_{\ell=1}^{\infty} \lambda_{\ell} B_{\ell}^{2}(q)\right\} = 2\sum_{\ell=1}^{\infty} \lambda_{\ell} q^2(1-q)^2.
\end{equation}
Based on Equations (\ref{equ: null_mean}) and (\ref{equ: null_var}), the mean and variance are increasing before 0.5 and decreasing after, with the peak at 0.5 and both ends equalling zero.
\subsection{Under $H_A$}
Under $H_A$, the observations follow $E(X_1)=\dots=E(X_{k^*})=\mu$ and $E(X_{k^*+1})=\dots=E(X_T)=\mu+\delta$.
\paragraph{a. Before the changepoint:}
When $k \leq k^*$,
\begin{equation}\nonumber
\begin{aligned}
S_{T, k}^A &= \frac{1}{\surd{T}}\left(\sum_{t=1}^k X_{t}-\frac{k}{T}\sum_{t=1}^{T} X_{t}\right)\\
&= \frac{1}{\surd{T}}\left[\sum_{t=1}^k(\mu+\epsilon_t)-\frac{k}{T}\left\{T\mu+(T-k^*)\delta+\sum_{t=1}
^T\epsilon_t\right\}\right]\\
&= \frac{1}{\surd{T}}\left(\sum_{t=1}^k\epsilon_t-\frac{k}{T}\sum_{t=1}^{T} \epsilon_{t}\right)-\frac{k}{\surd{T}} \frac{T-k^*}{T} \delta.
\end{aligned}
\end{equation}
Recall the definition of $S_{T}^0(q,u)$ and we denote $\frac{k}{\surd{T}} \frac{T-k^*}{T} \delta$ as $\delta_0$, so $S_{T, k}^A=S_{T, Tq}^A=S_{T}^0(q,u)-\delta_0$ and
\begin{equation} \nonumber
\begin{aligned}
    \left\|S_{T, k}^A\right\|&= \left\|S_{T}^0(q,u)-\delta_0\right\|=\left\|S_{T}^0(q,u)-\Gamma_{T}^0(q,u)+\Gamma_{T}^0(q,u)-\delta_0\right\|\\
    &\leq \left\|S_{T}^0(q,u)-\Gamma_{T}^0(q,u)\right\|+\left\|\Gamma_{T}^0(q,u)-\delta_0\right\|.
\end{aligned}
\end{equation}
Combining with Equation (\ref{equ: S0&Gamma0}), this implies $\left\|S_{T, k}^A\right\| \leq \left\|\Gamma_{T}^0(q,u)-\delta_0\right\|+o_p(1)$. On the other hand, we have
\begin{equation} \nonumber
\begin{aligned}
    \left\|\Gamma_{T}^0(q,u)-\delta_0\right\|&=\left\|\Gamma_{T}^0(q,u)-S_{T}^0(q,u)+S_{T}^0(q,u)-\delta_0\right\|\\
    & \leq \left\|\Gamma_{T}^0(q,u)-S_{T}^0(q,u)\right\|+\left\|S_{T}^0(q,u)-\delta_0\right\|.
\end{aligned}
\end{equation}
Again combining with Equation (\ref{equ: S0&Gamma0}), this implies $\left\|\Gamma_{T}^0(q,t)-\delta_0\right\| \leq \left\|S_{T, k}^A\right\|+o_p(1)$.
So $\left|\left\|S_{T, k}^A\right\| - \left\|\Gamma_{T}^0(q,u)-\delta_0\right\|\right| \leq o_p(1)$.
To simplify the notation, We write $Z_{T, k}$ for $\left\|\Gamma_{T}^0(q,u)-\delta_0\right\|^2$ and $Y_{T,k}$ for $\left\|S_{T, k}^A\right\|^2$, then we have 
\begin{equation}
    \surd Y_{T,k} - \surd Z_{T,k} \overset{\mathcal{P}}{\to}  0.
\end{equation}
And
\begin{equation} \nonumber
\begin{aligned}
    \left\|\Gamma_{T}^0(q,u)-\delta_0\right\|^2 &\stackrel{\mathcal{D}}{=}\left\|\sum_{l=1}^\infty \lambda_l^{1/2} B_l(q) \phi_l(u)-\frac{k}{T^{1/2}} \frac{T-k^*}{T} \delta\right\|^2\\
    &\stackrel{\mathcal{D}}{=}\int_0^1\left\{\sum_{l=1}^\infty \lambda_l^{1/2}B_l(q)\phi_l(u)-\frac{k}{T^{1/2}}\frac{T-k^*}{T} \delta\right\}^2 d u\\
    &\stackrel{\mathcal{D}}{=}\sum_{l=1}^\infty \lambda_l B_l^2(q)+\frac{k^2}{T}\frac{(T-k^*)^2}{T^2}||\delta||^2-\frac{2k}{T^{1/2}} \frac{T-k^*}{T}\sum_{l=1}^\infty \lambda_l^{1/2}B_l(q)\int_0^1 \phi_l \delta d u,
\end{aligned}
\end{equation}
i.e.
\begin{equation}
\label{equ: 1}
Z_{T, k} \stackrel{\mathcal{D}}{=} \sum_{l=1}^\infty \lambda_l B_l^2(q)+\frac{k^2}{T}\frac{(T-k^*)^2}{T^2}||\delta||^2-\frac{2k}{T^{1/2}} \frac{T-k^*}{T}\sum_{l=1}^\infty \lambda_l^{1/2}B_l(q)\int_0^1 \phi_l \delta d u.
\end{equation}
Then we calculate the expectation and variance of this approximation distribution based on Equation (\ref{equ: 1}) to get the conclusion when this is an alternative location and $q\leq k^*/T$ as follows,
\begin{equation}\nonumber
\begin{aligned}
    E\left(Z_{T, k}\right)&=\sum_{l=1}^\infty \lambda_l E\{B_l^2(q)\}+\frac{k^2}{T}\frac{(T-k^*)^2}{T^2}||\delta||^2-\frac{2k}{T^{1/2}} \frac{T-k^*}{T}\sum_{l=1}^\infty\lambda_l^{1/2}E\{B_l(q)\}\int_0^1 \phi_l \delta d u\\
    &=\sum_{l=1}^\infty\lambda_l q(1-q)+T||\delta||^2q^2\left(1-\frac{k^*}{T}\right)^2,
\end{aligned}
\end{equation}
\begin{equation}\nonumber
\begin{aligned}
    \operatorname{var}\left(Z_{T, k}\right)&=\sum_{l=1}^\infty \lambda_l^2 \operatorname{var}\{B_l^2(q)\}-\frac{4k}{T^{1/2}} \frac{T-k^*}{T}\sum_{m=1}^\infty\sum_{l=1}^\infty\lambda_m \lambda_l^{1/2}\operatorname{cov}\{B_m^2(q),B_l(q)\}\int_0^1 \phi_l \delta d u\\
    &\ \ \ \ \ +4\sum_{l=1}^\infty\lambda_l \left\{\int_0^1\phi_l(u)\delta (u)d u\right\}^2\frac{k^2(T-k^*)^2}{T^3}\operatorname{var}\{B_l(q)\}
    \\
    &=\sum_{l=1}^\infty 2\lambda_l^2 q^2(1-q)^2+4\sum_{l=1}^\infty\lambda_l \left\{\int_0^1\phi_l(u)\delta (u)d u\right\}^2\frac{k^2(T-k^*)^2}{T^3}q(1-q)\\
    &=\sum_{l=1}^\infty 2\lambda_l^2 q^2(1-q)^2+4\sum_{l=1}^\infty\lambda_l \left\{\int_0^1\phi_l(u)\delta(u)d u\right\}^2T\left( 1-\frac{k^*}{T}\right)^2 q^3(1-q).
\end{aligned}
\end{equation}
More concisely,
\begin{equation}
\label{equ: mean_before}
    E(Z_{T, k}) =
     \sum_{l=1}^\infty\lambda_l q\left(1-q\right)+T||\delta||^2q^2\left(1-\frac{k^*}{T}\right)^2,\ if\ q\leq \frac{k^*}{T},
\end{equation}
\begin{equation}
\label{equ: var_before}
    \operatorname{var}(Z_{T, k}) =
     a q^2\left(1-q\right)^2+b T \left(1-\frac{k^*}{T}\right)^2q^3\left(1-q\right),\ if\ q\leq \frac{k^*}{T},
\end{equation}
where $q=\frac{k}{T}$,
    $a=2\sum_{l=1}^\infty\lambda_l^2$, $b=4\sum_{l=1}^\infty \left\{ \int_0^1 \phi_l(u)\delta(u) du \right\}^2$.
\paragraph{b. After the changepoint:}   
Similarly, we deal with the case when $k > k^*$. Now
\begin{equation}\nonumber
\begin{aligned}
S_{T, k}^A &= \frac{1}{\surd{T}}\left(\sum_{t=1}^k X_{t}-\frac{k}{T}\sum_{t=1}^{T} X_{t}\right)\\
&= \frac{1}{\surd{T}}\left[\left\{k\mu+(k-k^*)\delta+\sum_{t=1}^k\epsilon_t\right\}-\frac{k}{T}\left\{T\mu+(T-k^*)\delta+\sum_{t=1}
^T\epsilon_t\right\}\right]\\
&= \frac{1}{\surd{T}}\left(\sum_{t=1}^k\epsilon_t-\frac{k}{T}\sum_{t=1}^{T} \epsilon_{t}\right)-\frac{k^*}{\surd{T}} \frac{T-k}{T} \delta.
\end{aligned}
\end{equation}
Following the procedure as before, we use the definition of $S_{T}^0(q,u)$ and write $\frac{k^*}{\surd{T}} \frac{T-k}{T} \delta$ as $\delta_1$, so $S_{T, k}^A=S_{T, Tq}^A=S_{T}^0(q,u)-\delta_1$.
By replacing $\delta_0$ by $\delta_1$ in the preceding calculations, we obtain a similar result and can, again, use $Z_{T, k}$ to show that under $H_A$, when $q>k^*/T$,
\begin{equation}
\label{equ: mean_after}
    E(Z_{T, k}) =
     \sum_{l=1}^\infty\lambda_l q\left(1-q\right)+T||\delta||^2\left(\frac{k^*}{T}\right)^2\left(1-q\right)^2,\ if\ q> \frac{k^*}{T},
\end{equation}
\begin{equation}
\label{equ: var_after}
    \operatorname{var}(Z_{T, k}) =
     a q^2\left(1-q\right)^2+b T \left(\frac{k^*}{T}\right)^2q\left(1-q\right)^3,\ if\ q> \frac{k^*}{T},
\end{equation}
where $q=\frac{k}{T}$,
    $a=2\sum_{l=1}^\infty\lambda_l^2$, $b=4\sum_{l=1}^\infty \left\{ \int_0^1 \phi_l(u)\delta(u) du \right\}^2$.
    
Based on (\ref{equ: mean_before}) and (\ref{equ: mean_after}), when $T$ is large, either before or after the changepoint, the form of the mean will be dominated by the $T$ dependent terms. Thus, the mean is approximately increasing before the changepoint and approximately decreasing after, with the peak exactly on the changepoint location and both ends equalling zero. Because the theoretical form of the mean is complicated, we use a two-piece piecewise linear model with fixed ends to approximate it. 

\begin{figure}[H]
    \centering
    \includegraphics[width=0.8\textwidth]{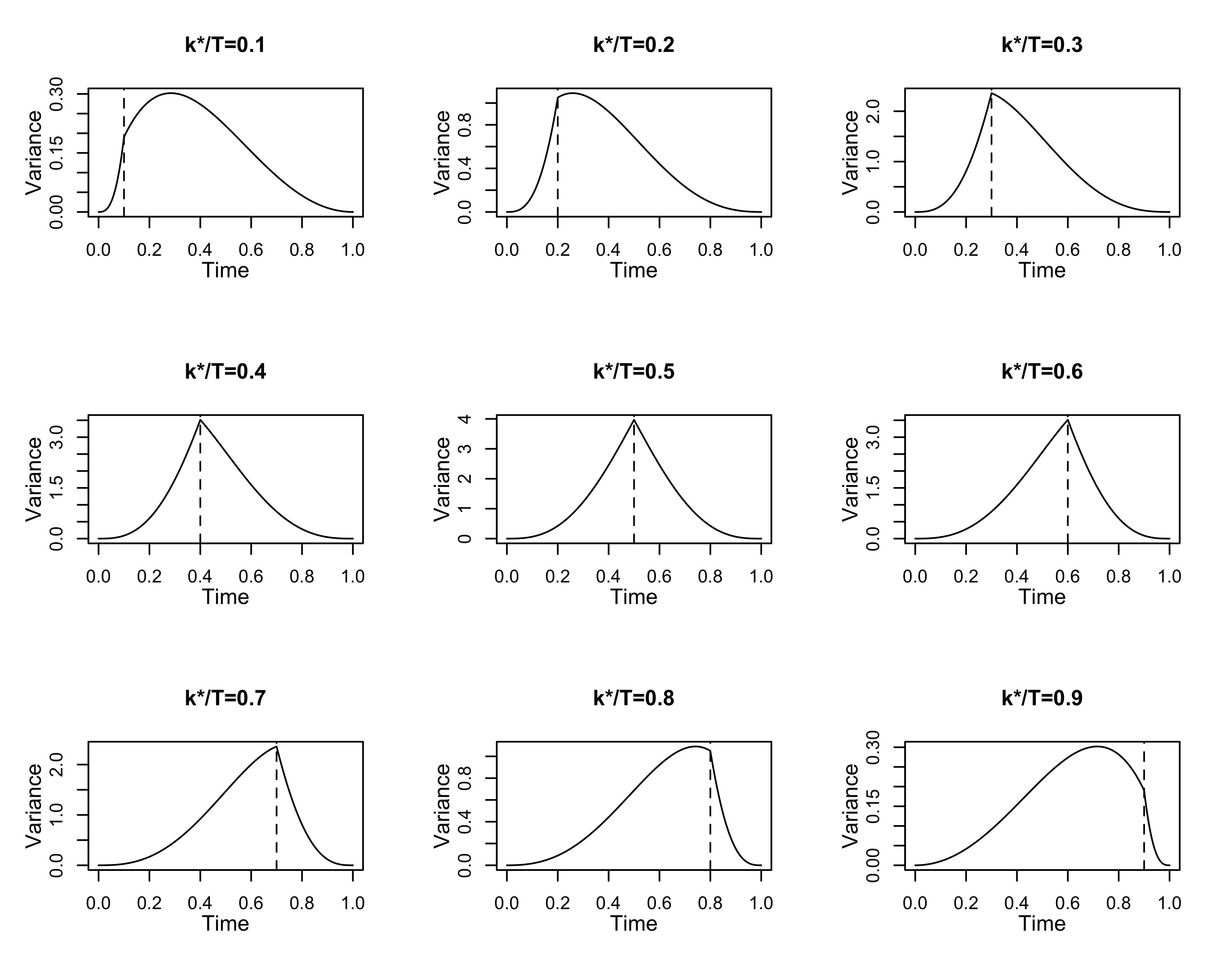}
    \caption{Proposed variance approximation for $Y_{T,k}$ process in Equation \ref{eq: HA_var} with changepoint at different locations when $a=1$, $b=5$, $T=50$.}
    \label{fig: th_var}
\end{figure}

We keep the special form of variance to better capture the uncertainty of the $Y_{T,k}$ process. To have a better visualization of the proposed theoretical form for variance, we plot the variance under cases with different changepoint as shown in Figure \ref{fig: th_var}. From this, we can see that the variance is always relatively large in the middle no matter where the changepoint is.

\section{Properties of $Y_{T,k}$ Process}
\label{app: Y Process}
\paragraph{Approximation of $Y_{T,k}$ Process}
To illustrate how well the approximation form can mimic the $Y_{T, k}(\bs)$ process when a changepoint exists at location $\bs$, $T$ curves are simulated and we consider changepoint 0.6 on the scaled time domain [0, 1], i.e. if $T=$100, the changepoint is 60. The noise observations and change functions are generated following the same formulas as those in the simulation. The magnitude of the change function $\rho$ is tuned for different scenarios to maintain the target signal-to-noise ratio (SNR), which is detailed introduced in Section \ref{sec: datageneration}. For each setting, the simulation is repeated 500 times. We collect $Y_{T,k}$ at each time point and then calculate the mean and variance of all 500 simulated $Y_{T,k}$'s. In addition, we estimate the eigenvalues and eigenfunctions to obtain the proposed theoretical approximation $E\{Z_{T, k}(\bs)\}$ and $\operatorname{var}\{Z_{T, k}(\bs)\}$ as in \eqref{eq: HA_mean} and \eqref{eq: HA_var}. The estimation procedures follow those from \cite{aue2018detecting}. Figure \ref{fig: mean_var_pairs} compares the mean and variance from the simulations and those derived based on the theoretical approximation. The first row in the figure is the result for mean and the second for variance. It is seen that when $T=$50 and 100, the approximation captures the trend and uncertainty of the $Y_{T,k}$ process very well.
\paragraph{Piecewise Linear Fit of the mean of $Y_{T,k}$ Process}
 To present how we use the piecewise linear model to model $Y_{T,k}$ process, we consider the case with $T=50$ time points and changepoint at 0.7 and show the results in Figure \ref{fig: plm_approx}. The noise functional data and change function ($\rho=0.8$) are generated according to the procedure in the simulation. We pick two interesting simulations: one has $Y_{T,k}$ process with a peak right at the changepoint, and the other one has a peak slightly off. For each $Y_{T,k}$ process, to get the best two-piece piecewise linear fit with two fixed ends, we use grid search for the slope and break point and pick the model with the smallest residual sum of squares, which is shown as the dashed line in the figure.

\section{Markov Chain Monte Carlo algorithm}
\label{app: mcmc}
We denote the entire $Y_{T,k}$ process for all $N$ locations as
$$\bfyy=\left(Y_{T,1}(\mathbf{s}_1),\ldots,\ Y_{T,1}(\mathbf{s}_N),\ Y_{T,2}(\mathbf{s}_1),\dots,Y_{T,2}(\mathbf{s}_N),\ldots,Y_{T,T-1}(\mathbf{s}_1),\dots, Y_{T,T-1}(\mathbf{s}_N)\right)^T.$$
To make the symbols concise, we use $\bbeta^0$, $\bc^0$, $a^0$, and $\bb^0$ to represent $log(-\bbeta)$, $\Phi^{-1}(\bc)$, $log(a)$, and $log(\bb)$. Then we can get the joint likelihood of all the parameters conditioning on the observations
\begin{equation}\nonumber
\begin{aligned}
    &f(\bbeta^0, a^0, \bb^0, \bc^0, \bmu_\beta, \mu_a, \bmu_b,\bmu_c, \sigma^2_\beta, \sigma^2_a, \sigma^2_b, \sigma^2_c, \phi, \phi_s, \phi_t|\bfyy)\\
    \propto &f(\bfyy|\bbeta^0, a^0, \bb^0, \bc^0, \phi_s, \phi_t)f(\bbeta^0|\bmu_\beta,\sigma^2_\beta,\phi)f(a^0|\mu_a, \sigma^2_a)f(\bb^0|\bmu_b,\sigma^2_b,\phi)f(\bc^0|\bmu_c,\sigma^2_c,\phi)\\
    &f(\bmu_\beta)f(\mu_a)f(\bmu_b)f(\bmu_c)f(\sigma^2_\beta)f(\sigma^2_a)f(\sigma^2_b)f(\sigma^2_c)f(\phi)f(\phi_s)f(\phi_t).
\end{aligned}
\end{equation}
Further, we can have the fully conditional likelihood for each parameter as follows,
$$
\begin{aligned}
    &f(\bmu_\beta|\cdot) \propto f(\bbeta^0|\bmu_\beta,\sigma^2_\beta,\phi)f(\bmu_\beta),\\
    & f(\mu_a|\cdot) \propto f(a^0|\mu_a, \sigma^2_a)f(\mu_a),\\
    & f(\bmu_b|\cdot) \propto f(\bb^0|\bmu_b,\sigma^2_b,\phi)f(\bmu_b),\\
    &f(\bmu_c|\cdot) \propto f(\bc^0|\bmu_c,\sigma^2_c,\phi)f(\bmu_c),\\
    &f(\bbeta^0|\cdot) \propto f(\bfyy|\bbeta^0, a^0, \bb^0, \bc^0, \phi_s, \phi_t)f(\bbeta^0|\bmu_\beta,\sigma^2_\beta,\phi),\\
    &f(a^0|\cdot) \propto f(\bfyy|\bbeta^0, a^0, \bb^0, \bc^0, \phi_s, \phi_t)f(a^0|\mu_a, \sigma^2_a),\\
    & f(\bb^0|\cdot) \propto f(\bfyy|\bbeta^0, a^0, \bb^0, \bc^0, \phi_s, \phi_t)f(\bb^0|\bmu_b, \sigma^2_b,\phi),\\
    &f(\bc^0|\cdot) \propto f(\bfyy|\bbeta^0, a^0, \bb^0, \bc^0, \phi_s, \phi_t)f(\bc^0|\bmu_c, \sigma^2_c,\phi),\\
    &f(\phi|\cdot) \propto f(\bbeta^0|\bmu_\beta,\sigma^2_\beta,\phi)f(\bb^0|\bmu_b, \sigma^2_b,\phi)f(\bc^0|\bmu_c,\sigma^2_c,\phi)f(\phi),\\
    &f(\phi_s|\cdot)\propto f(\bfyy|\bbeta^0, a^0, \bb^0, \bc^0, \phi_s, \phi_t)f(\phi_s),\\
    &f(\phi_t|\cdot)\propto f(\bfyy|\bbeta^0, a^0, \bb^0, \bc^0, \phi_s, \phi_t)f(\phi_t),\\
    & f(\sigma_\beta^2|\cdot)\propto f(\bbeta^0|\bmu_\beta,\sigma^2_\beta,\phi)f(\sigma_\beta^2),\\
    & f(\sigma_a^2|\cdot)\propto f(a^0|\mu_a,\sigma^2_a)f(\sigma_a^2),\\
    & f(\sigma_b^2|\cdot)\propto f(\bb^0|\bmu_b,\sigma^2_b,\phi)f(\sigma_b^2),\\
    & f(\sigma_c^2|\cdot)\propto f(\bc^0|\bmu_c,\sigma^2_c,\phi)f(\sigma_c^2).
\end{aligned}
$$
For the variance parameters $\sigma^2_i, i=\beta, a, b, c$, we use Gibbs sampling and we use Metropolis-Hasting-within-Gibbs to update other parameters. 

\paragraph{a. Variance parameters}
To get the posterior distribution for $\sigma^2_i, i=\beta, a, b, c$, we first derive the posterior distribution of the variance parameter with conjugate prior in the general case. Recall the multivariate normal distribution for an $n \times 1$ vector $\bfyy$:
$$
f(\bfyy|\bmu,\sigma^2\bSigma)=\left\{(2\pi)^n|\sigma^2\bSigma|\right\}^{-1/2}\exp\left\{-\frac{1}{2\sigma^2}(\bfyy-\bmu)^T\bSigma^{-1}(\bfyy-\bmu)\right\}.
$$
Assume that $\sigma^2 \sim IG(\alpha_1,\alpha_2)$, i.e. $f(\sigma^2)\propto (\sigma^2)^{-(\alpha_1+1)}\exp\left\{-\frac{\alpha_2}{\sigma^2}\right\}$. Then
$$
\begin{aligned}
f(\sigma^2|\cdot)&\propto f(\bfyy|\bmu,\sigma^2\bSigma)f(\sigma^2)\\
&\propto (\sigma^2)^{-n/2}(\sigma^2)^{-(\alpha_1+1)}\exp\left\{-\frac{1}{2\sigma^2}(\bfyy-\bmu)^T\bSigma^{-1}(\bfyy-\bmu)\right\}\exp\left\{-\frac{\alpha_2}{\sigma^2}\right\}\\
&\propto (\sigma^2)^{-(n/2+\alpha_1+1)}\exp\left[-\frac{1}{\sigma^2}\left\{\frac{(\bfyy-\bmu)^T\bSigma^{-1}(\bfyy-\bmu)}{2}+\alpha_2\right\}\right],
\end{aligned}$$
which means $\sigma^2|\cdot\sim IG\left(n/2+\alpha_1,(\bfyy-\bmu)^T\bSigma^{-1}(\bfyy-\bmu)/2+\alpha_2\right)$.

Now we apply the conclusion above to our cases. From Stage III in our model, $\sigma^2_i \sim IG(0.1,0.1),\ i = \beta, c, a, b$, so $\alpha_1=0.1$, $\alpha_2=0.1$. Following $f(\sigma_\beta^2|\cdot)\propto f(\bbeta^0|\bmu_\beta,\sigma^2_\beta,\phi)f(\sigma_\beta^2)$, we have 
$$\sigma_\beta^2|\cdot\sim IG\left(\frac{N}{2}+\alpha_1,\frac{(\bbeta^0-\bmu_\beta)^T\bSigma(\phi)^{-1}(\bbeta^0-\bmu_\beta)}{2}+\alpha_2\right).$$
Similarly, we can get
$$
\begin{aligned}
\sigma_a^2|\cdot &\sim IG\left(\frac{1}{2}+\alpha_1,\frac{(a^0-\mu_a)^2}{2}+\alpha_2\right),\\
\sigma_b^2|\cdot &\sim IG\left(\frac{N}{2}+\alpha_1,\frac{(\bb^0-\bmu_b)^T\bSigma(\phi)^{-1}(\bb^0-\bmu_b)}{2}+\alpha_2\right),\\
\sigma_c^2|\cdot &\sim IG\left(\frac{N}{2}+\alpha_1,\frac{(\bc^0-\bmu_c)^T\bSigma(\phi)^{-1}(\bc^0-\bmu_c)}{2}+\alpha_2\right).\\
\end{aligned}
$$

\paragraph{b. \boldsymbol{$\bbeta^0$}, $a^0$, \boldsymbol{$\bb^0$}, \boldsymbol{$\bc^0$}, \boldsymbol{$\bmu_\beta$}, $\mu_a$, $\bmu_b$, \boldsymbol{$\bmu_c$}}
For the transformed parameters $\bbeta^0$, $a^0$, $\bb^0$, $\bc^0$ and the mean parameters $\bmu_\beta$, $\mu_a$, $\bmu_b$, $\bmu_c$ in Stage II, we use Metropolis-Hasting-within-Gibbs with symmetric proposal distribution. 

In the following, we take the parameter $\bc^0$ as an example to illustrate the process in Iteration $j+1$. We use $\bc^{0(j)}$ and $\bc^{0*}$ to represent the sample in $j$th iteration and the new proposed sample. And we pick $N(\bc^{0(j)}, \sigma^2)$ as the proposal distribution, where $\sigma^2$ is a tuning parameter.
\begin{enumerate}[label=(\roman*)]
    \item  Generate a random candidate state $\bc^{0*} \sim N(\bc^{0(j)}, \sigma^2)$.
    \item Calculate the acceptance rate $A(\bc^{0*}, \bc^{0(j)})=min \left\{1, \frac{f(\bc^{0*}|\cdot)T(\bc^{0(j)}|\bc^{0*})}{f(\bc^{0(j)}|\cdot)T(\bc^{0*}|\bc^{0(j)})}\right\}$.
    \item Then generate a random number $u \in [0,1]$ from the uniform distribution on [0, 1]. If $u \leq A(\bc^{0*}, \bc^{0(j)})$, we accept the new state and set $\bc^{0(j+1)}=\bc^{0*}$. Otherwise, we reject the new state and set $\bc^{0(j+1)}=\bc^{0(j)}$.
\end{enumerate}

Note that because of the symmetry of the proposal distribution and supposing in the $j$th iteration, we have updated all the other parameters except the three range parameters $\phi$, $\phi_s$, $\phi_t$ and the variance parameters $\sigma^2_\beta$, $\sigma^2_a$, $\sigma^2_b$, $\sigma^2_c$ then
$$\frac{f(\bc^{0*}|\cdot)T(\bc^{0(j)}|\bc^{0*})}{f(\bc^{0(j)}|\cdot)T(\bc^{0*}|\bc^{0(j)})}=\frac{f(\bfyy|\bbeta^{0(j+1)}, a^{0(j+1)}, \bb^{0(j+1)}, \bc^{0*}, \phi_s^{(j)}, \phi_t^{(j)})f(\bc^{0*}|\bmu_c^{(j+1)}, \sigma^{2(j)}_c,\phi^{(j)})}{f(\bfyy|\bbeta^{0(j+1)}, a^{0(j+1)}, \bb^{0(j+1)}, \bc^{0(j)}, \phi_s^{(j)}, \phi_t^{(j)})f(\bc^{0(j)}|\bmu_c^{(j+1)}, \sigma^{2(j)}_c,\phi^{(j)})}.$$

\paragraph{c. Range parameters}
For the range parameters, we still use a normal random walk as the proposal distribution, but note that the range parameters should be positive. For example, to get posterior samples for $\phi$, the acceptance rate is $A(\phi^*, \phi^{(j)}) = min\left\{1, \frac{f(\phi^*|\cdot)T(\phi^{(j)}|\phi^*)}{f(\phi^{(j)}|\cdot)T(\phi^*|\phi^{(j)})}\right\}$. The random candidate $\phi^*$ is generated from $N(\phi^{(j)},\sigma^2)$ and $\phi^*$ is positive, where $\sigma^2$ is a tuning parameter and can be different from that we use to generate $\bc^{0*}$. Assume that in the $(j+1)$th iteration, $\phi$ is updated right after $\bc^0$, and the other two range parameters $\phi_s$, $\phi_t$ and the variance parameters are not updated yet. Based on the proposal distribution, we have

$$T(\phi^*|\phi^{(j)})\propto \frac{1}{\sigma}\frac{\bphi\left(\frac{\phi^*-\phi^{(j)}}{\sigma}\right)}{1-\bPhi\left(\frac{-\phi^{(j)}}{\sigma}\right)},$$
and
$$
\begin{aligned}
 &\frac{f(\phi^*|\cdot)T(\phi^{(j)}|\phi^*)}{f(\phi^{(j)}|\cdot)T(\phi^*|\phi^{(j)})}\\
=& \frac{f(\bbeta^{0(j+1)}|\bmu_\beta^{(j+1)},\sigma^{2(j)}_\beta,\phi^*)f(\bb^{0(j+1)}|\bmu_b^{(j+1)}, \sigma^{2(j)}_b,\phi^{*})f(\bc^{0(j+1)}|\bmu_c^{(j+1)},\sigma^{2(j)}_c,\phi^*)f(\phi^*)}{f(\bbeta^{0(j+1)}|\bmu_\beta^{(j+1)},\sigma^{2(j)}_\beta,\phi^{(j)})f(\bb^{0(j+1)}|\bmu_b^{(j+1)}, \sigma^{2(j)}_b,\phi^{(j)})f(\bc^{0(j+1)}|\bmu_c^{(j+1)},\sigma^{2(j)}_c,\phi^{(j)})f(\phi^{(j)})}\times\\
&\frac{\bphi\left(\frac{\phi^{(j)} - \phi^*}{\sigma}\right)\left\{1-\bPhi\left(\frac{-\phi^{(j)}}{\sigma}\right)\right\}}{\bphi\left(\frac{\phi^* - \phi^{(j)}}{\sigma}\right)\left\{1-\bPhi\left(\frac{-\phi^*}{\sigma}\right)\right\}},
\end{aligned}
$$
where $\bphi$, $\bPhi$ are the probability density function and cumulative density function of the standard normal distribution.
To ensure the convergence of the chains, we try several sets of different initial values for all parameters and evaluate the difference between those chains by Gelman–Rubin diagnostic. Geweke’s diagnostic is applied to determine the burn‐in period.
\section{Data Generation}
\label{app: appendix data generation}
With Fourier basis functions, we generate the functional data according to
$$
\begin{aligned}
&\varepsilon_{\bs, t} = \sum_{l=1}^L \xi _{\bs,t}^l \nu_l,\ \bs \in \md_R, \  t=1,\dots,T, \\
&\delta_\bs = \sum_{l=1}^L \eta^{l}_\bs \nu_l,\ \bs \in 
\md_a,
\end{aligned}
$$ 
where $\varepsilon_{\bs, t}$ is the independent curve at location $\bs$ at time point $t$ and $\delta_\bs$ is the change function for the alternative location.
To make the symbols consistent with that in our R code, here $\nu_l$ is the $l$th basis function in R. To make the index of coefficients easier to understand, we rewrite the above two equations as follows,
 $$\varepsilon_{\bs, t}(u) = A_{t,0}(\bs) + \sum_{l=1}^{(L-1)/2} \left\{\surd{2}A_{t,l}(\bs) cos(2\pi lu) +\surd{2}B_{t,l}(\bs) sin(2\pi lu)\right\},$$
 $$\delta_\bs(u) = \widetilde{A}_{0}(\bs) + \sum_{l=1}^{(L-1)/2} \left\{\surd{2}\widetilde{A}_{l}(\bs)cos(2\pi lu) +\surd{2}\widetilde{B}_{l}(\bs) sin(2\pi lu)\right\}.$$
Note that for the cosine based basis functions, we use $A$ and $\widetilde{A}$ to denote its coefficient. And $B$ and $\widetilde{B}$ are for sine based basis functions. When programming in R, $\surd{2}$ is multiplied in front of the basis to make sure the norm of each basis is 1. Similar results can be easily obtained by multiplying some constant if another programming language is used.

To guarantee the smoothness of the functional data with and without change function, we expect the coefficients decay with the frequency of the basis function. And the sine and cosine basis would not influence the magnitude of coefficients if they share the same frequency. So we divide the coefficients into several groups and coefficients in the same group will share the same fluctuation.

Recall the smoothness and decay properties of Fourier coefficients: A piecewise continuous function has Fourier coefficients that decay as $1/n$. And also a conclusion in stochastic sequence: If $\left(X_{n}\right)$ is a stochastic sequence such that each element has finite variance, then
    $$
X_{n}-E\left(X_{n}\right)=O_{p}\left[\{\operatorname{var}\left(X_{n}\right)\}^{1/2}\right].
$$
Moreover, if $a_{n}^{-2} \operatorname{var}\left(X_{n}\right)=\operatorname{var}\left(a_{n}^{-1} X_{n}\right)$ is a null sequence for a sequence $\left(a_{n}\right)$ of real numbers, then $a_{n}^{-1}\left\{X_{n}-E\left(X_{n}\right)\right\}$ converges to zero in probability by Chebyshev's inequality, so
$$
X_{n}-E\left(X_{n}\right)=o_{p}\left(a_{n}\right).
$$
To make sure that the independent curves, change functions, and the final observations are at least piecewise continuous, the coefficients should satisfy that for any $\bs$,
$$A_{t,l}(\bs)=o_p\left(\frac{1}{l}\right),\  B_{t,l}(\bs)=o_p\left(\frac{1}{l}\right),$$
$$\widetilde{A}_{l}(\bs)=o_p\left(\frac{1}{l}\right),\  \widetilde{B}_{l}(\bs)=o_p\left(\frac{1}{l}\right),$$
$$\widetilde{A}_{l}(\bs)+A_{t,l}(\bs)=o_p\left(\frac{1}{l}\right),\  \widetilde{B}_{l}(\bs)+B_{t,l}(\bs)=o_p\left(\frac{1}{l}\right),\ t=k_i^*+1,\dots,T.$$

If $A_{n,l}(\bs)$, $B_{t,l}(\bs)\sim N(0, \frac{r_1}{l^3})$ and $\widetilde{A}_{l}(\bs)$, $\widetilde{B}_{l}(\bs)\sim N(\rho \frac{1}{l^2}, \frac{r_2}{l^3})$, the conditions above can be guaranteed, where $r_1$, $r_2$ and $\rho$ are some constants. We choose $r_1=\frac{1}{2}$ and $r_2=\frac{1}{10}$, so the range of independent curves is well controlled and the change functions for all locations share a similar shape, which is a common situation in the spatial correlated real dataset. And the spatial correlation structure is the same as that we use to generate changepoints.

\section{Initial Values}
For each location $\bs$, we get the estimates for parameters $\beta(\bs)$, $c(\bs)$, $a(\bs)$ and $b(\bs)$ and denote them as $\hat{\beta}(\bs)$, $\hat{c}(\bs)$, $\hat{a}(\bs)$ and $\hat{b}(\bs)$. To start the MCMC, we assign initial values for all parameters as follows. 
$$
\begin{aligned}
&\bmu_{\beta, 0} = \frac{1}{N} \sum_{n=1}^N log\left\{-\hat{\beta}(\bs_n)\right\}\textbf{1}_N,\\
&\bmu_{c, 0} = \textbf{0}_N,\\
&\mu_{a, 0} = \frac{1}{N} \sum_{n=1}^N log\left\{\hat{a}(\bs_n)\right\},\\
&\bmu_{b, 0} = \left(log\left\{\hat{b}(\bs_1)\right\}, \ldots, log\left\{\hat{b}(\bs_N)\right\}\right),\\
&\bbeta_0 = \left(log\left\{-\hat{\beta}(\bs_1)\right\}, \ldots, log\left\{-\hat{\beta}(\bs_N)\right\}\right),\\
&\bc_0 = \left(\bPhi^{-1}\left\{\hat{c}(\bs_1)\right\}, \ldots, \bPhi^{-1}\left\{\hat{c}(\bs_N)\right\}\right),\\
&a_0 = \frac{1}{N} \sum_{n=1}^N log\left\{\hat{a}(\bs_n)\right\},\\
&\bb_0 = \left(log\left\{\hat{b}(\bs_1)\right\}, \ldots, log\left\{\hat{b}(\bs_N)\right\}\right),\\
&\sigma_\beta^2 = 1, \sigma_c^2 = 1, \sigma^2_a = 0.5, \sigma_b^2 = 1, \phi_s = 2, \phi_t = 0.2,\\
&\phi = 5\text{ (the real }\phi\text{ value if in the simulation)}.
\end{aligned}
$$
In the following, we illustrate the way to get the estimates for one location, so we ignore the index for the location and denote the estimates at one location as $\hat{\beta}$, $\hat{c}$, $\hat{a}$, and $\hat{b}$.

For each location, we use the package \texttt{fChange} which implements the method introduced in \cite{aue2018detecting} to get the estimates of changepoint, eigenvalues, eigenfunctions and change function denoted as $\hat{c}$, $\hat{\lambda_l}$, $\hat{\psi_l}(u)$, and $\hat{\delta}(u)$. The estimate of the changepoint, $\hat{c}$, is the time when the CUSUM statistic gets the maximum value. With the changepoint estimated, functional data can be split into two parts and the difference of the mean functions from those two parts is the estimated change function. The eigenvalues and eigenfunctions are estimated based on the estimated long-run covariance operator in Appendix \ref{app: eigen of kernel}. Based on those and Theorem \ref{thm:m&v_HA}, we further get the estimated $a$ and $b$ as $\hat{a}=2\sum_{l=1}^\infty\hat{\lambda}_l^2$ and $\hat{b}=4\sum_{l=1}^\infty \left\{ \int_0^1 \hat{\psi}_l(u)\hat{\delta}(u) du \right\}^2$.
For $\hat{\beta}$, we generate a sequence of possible values, use the two-piece piecewise linear model with fixed ends $y = \beta\{(\hat{c}-1)q+(q-\hat{c})\mathds{1}(q\geq \hat{c})\},\ q=\frac{k}{T},\ k=1,\dots,T-1$ based on the changepoint estimate $\hat{c}$ from the FF method, and pick the $\beta$ that can model the $Y_{T,k}$ process with the minimum mean squared error. 

The estimators of the parameters need not necessarily be very accurate, since we just would like to provide reasonable initial values that are roughly on the same scale as that of the true values, which can help on the convergence of the chains.

\section{Additional Evidence for Simulation}
\label{app: simulation}
\begin{figure}[h]
  \centering
  \includegraphics[width=0.975\linewidth, right]{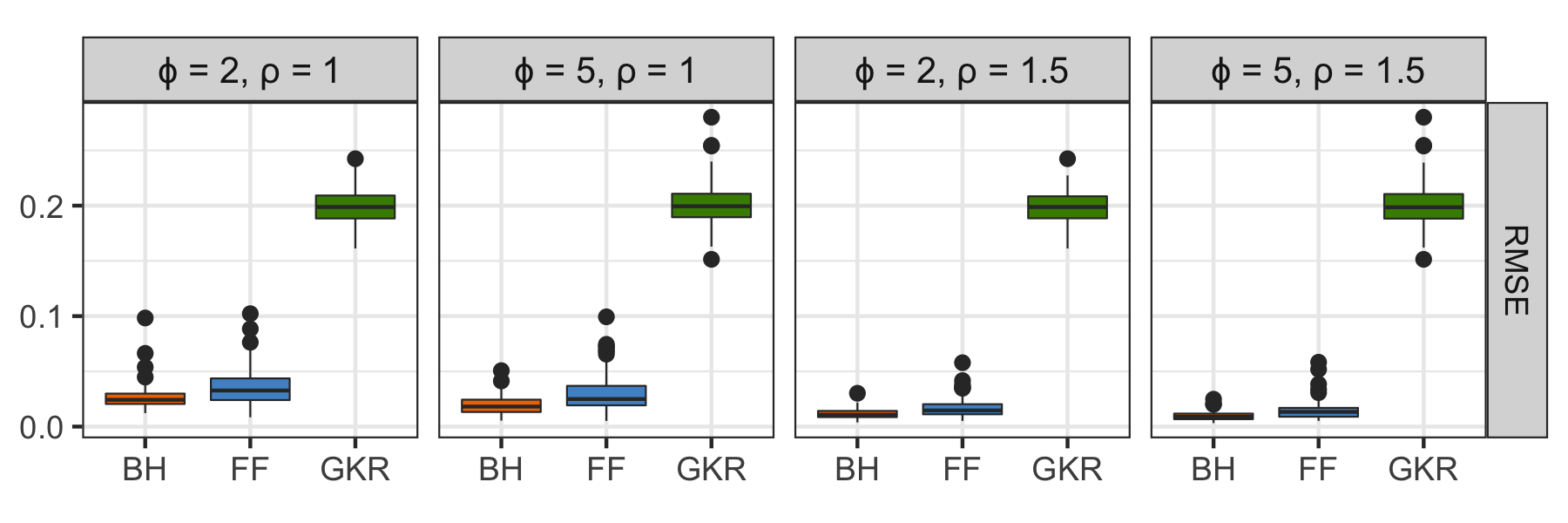}  
  \caption{Boxplots of RMSE from three different methods under four settings. "GKR" indicates the method in  \cite{gromenko2017detection}. The settings and other method labelling are the same as those in Figure \ref{fig: simulation_boxplot}.}
  \label{fig: sim_rmse_3}
\end{figure}

\paragraph{RMSE from the method in \cite{gromenko2017detection}} Figure \ref{fig: sim_rmse_3} shows the RMSE from all three methods. 
\paragraph{Further comparisons between BH and FF}
From Figure \ref{fig: simulation_boxplot}, it seems that the FF method also benefits from the stronger spatial correlation, which should not be, in theory. To further explore how BH and FF react to different spatial correlations given the same data generation seed number, we examine two types of pairwise differences where each pair shares the same seed number.   

\begin{figure}[h]
\begin{subfigure}{.5\textwidth}
  \centering
  \includegraphics[width=\linewidth]{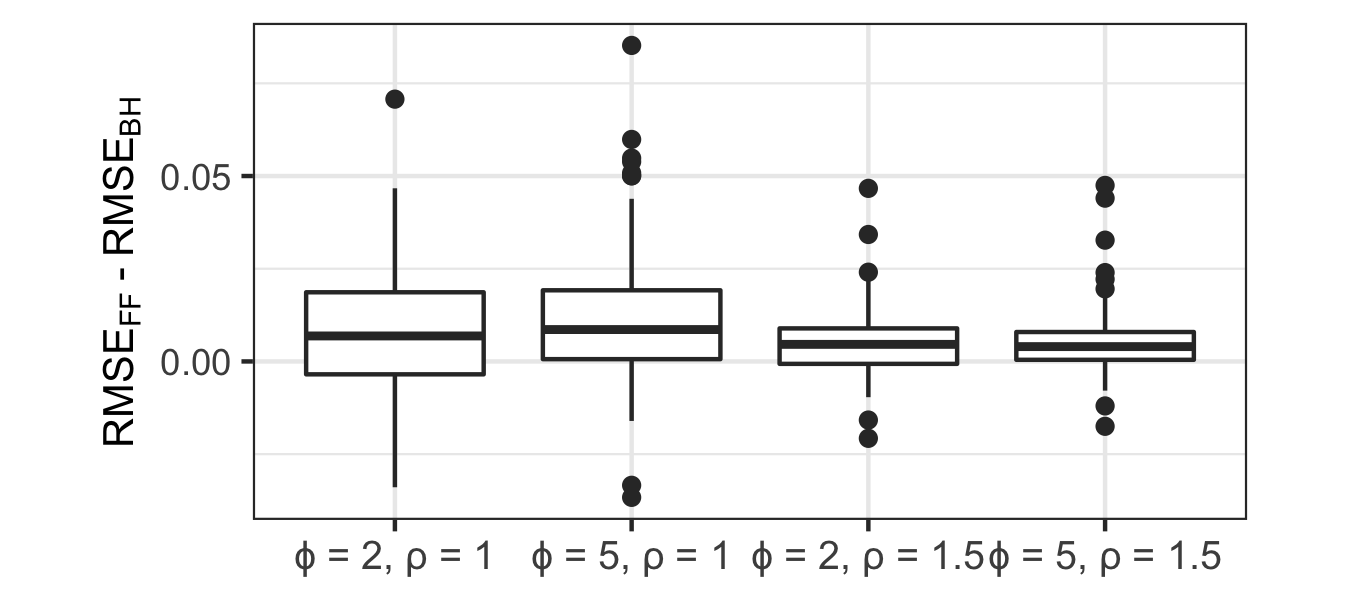}  
  \caption{}
  \label{fig: paircomp1}
\end{subfigure}
\begin{subfigure}{.5\textwidth}
  \centering
  \includegraphics[width=\linewidth]{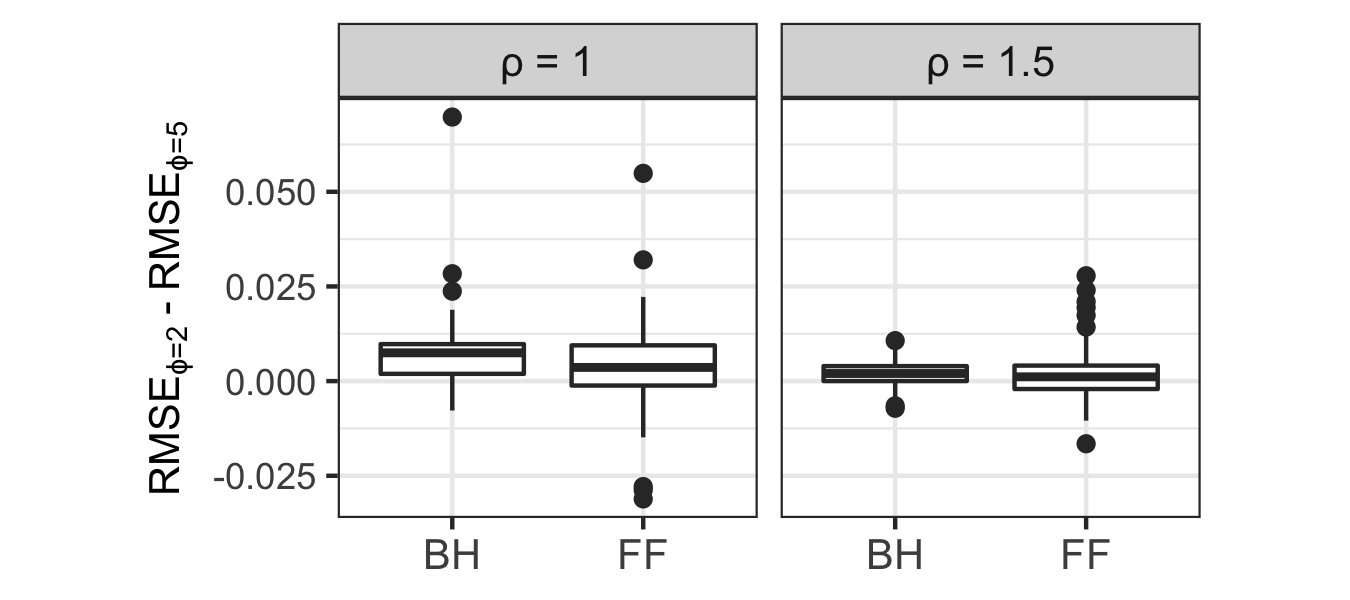}  
  \caption{}
  \label{fig: paircomp2}
\end{subfigure}
\caption{(a) Boxplots of pairwise differences between RMSE from FF and RMSE from BH  under four different settings. (b) Boxplots of pairwise differences between RMSE from $\phi=2$ and RMSE from $\phi=5$ for BH and FF under both weaker and stronger signal strength.}
\label{fig: add_evidence}
\end{figure}

We first take the pairwise RMSE difference between the FF and BH method for each parameter setting, as shown in Figure \ref{fig: paircomp1}. When spatial correlation is stronger ($\phi=5$), the RMSE reduction by using the BH appears more significant than that with the weaker spatial correlation ($\phi=2$).
Then we take the pairwise RMSE difference between $\phi=2$ and $\phi=5$ for each of the FF and BH methods and for both $\rho=1$ and $\rho=1.5$, as shown in Figure \ref{fig: paircomp2}. Again, the RMSE reduction of BH by having $\phi=5$ as opposed to $\phi=2$ for both $\rho=1$ and $\rho=1.5$ appears more significant than the RMSE reduction of FF due to a stronger spatial correlation. 
Both plots (a) and (b) show that the BH essentially benefits from the stronger spatial correlation while there is no clear evidence that FF enjoys strong spatial correlation.


\section{Real Data}
\begin{figure}[H]
    \centering
    \includegraphics[width=0.5\textwidth]{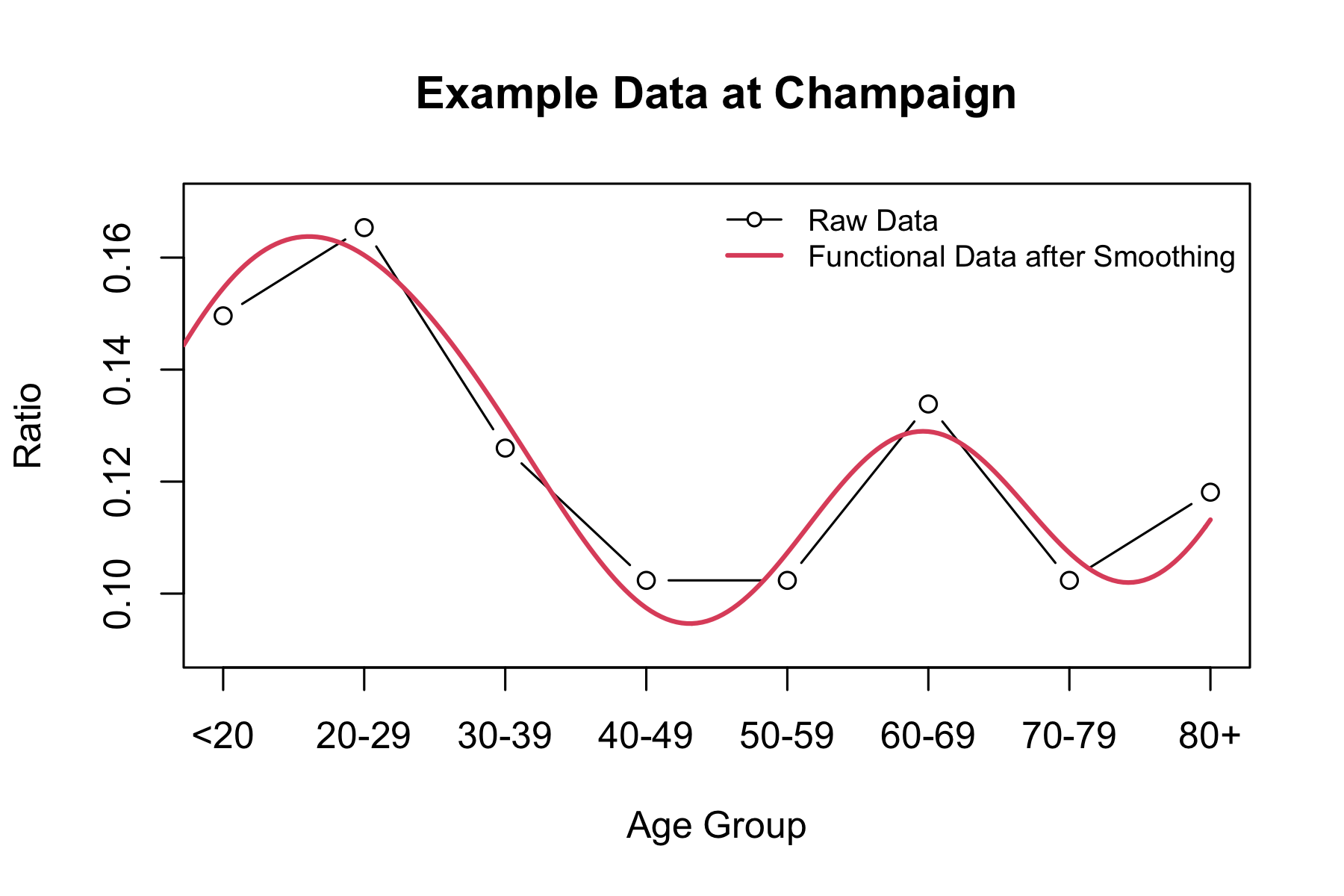}
    \caption{The ratio of cases in each age group in Champaign county on Jan. 5th, 2021 (black line with dots) and functional time series after smoothing (red). The x-axis labels the corresponding age groups evenly spread between 0 and 1.}
    \label{fig: covid_fdata}
\end{figure}

\textbf{COVID-19 Data in Illinois} We use 7 Fourier basis functions to smooth the raw data. An example about the raw data and the functional time series after smoothing is shown in Figure \ref{fig: covid_fdata}.
\label{app: realdata}
\end{document}